\documentclass[%
preprint,
 amsmath,amssymb,
 aps,
]{revtex4-1}

\usepackage{graphicx}
\usepackage{dcolumn}
\usepackage{bm}

\usepackage{amssymb,amsmath,amsthm,accents}
\usepackage{amscd,amsfonts}
\usepackage{fullpage}
\usepackage{subfig}
\usepackage{float}
\usepackage{lineno}
\usepackage{mathrsfs}
\usepackage{hyperref}
\usepackage{url}
\usepackage{color}
\usepackage{xcolor}
\usepackage{epsfig}
\usepackage{epstopdf}
\DeclareGraphicsExtensions{.eps,.pdf,.png,.jpg,.jpeg}
\usepackage{feynmp}
\begin{document}

\title{Dark energy via multi-Higgs doublet models:\\
	accelerated expansion of the Universe in inert doublet model scenario}
	\author{Muhammad Usman}
	\email{muhammad\_usman\_sharif@yahoo.com, \\ muhammadusman@sns.nust.edu.pk}
	\affiliation{
		School of Natural Sciences (SNS),\\
		National University of Sciences and Technology (NUST),\\
		Sector H-12, Islamabad 44000, Pakistan
	}
	
	\date{\today}
	\begin{abstract}
	Scalar fields are among the possible candidates for dark energy. This paper is devoted to the scalar fields from the inert doublet model, where instead of one as in the standard model, two SU(2) Higgs doublets are used. The component fields of one SU(2) doublet ($\phi_1$) act in an identical way to the standard model Higgs while the component fields of the second SU(2) doublet ($\phi_2$) are taken to be the dark energy candidate (which is done by assuming that the phase transition in the field has not yet occurred
	). It is found that one can arrange for late time acceleration (dark energy) by using an SU(2) Higgs doublet in the inert Higgs doublet model, whose vacuum expectation value is zero, in the quintessential regime.
	\end{abstract}
	\maketitle
\section{Introduction}
There are at least three components which contribute to the total energy density of the Universe. These are non-relativistic matter, relativistic matter and dark energy. Dark energy, which constitutes about $70\%$ of the present energy density of the Universe is one of the most discussed and hot topics of this era. Several possible candidates (explanations) of inflation and dark energy are discussed in the literature, a few of them are: the cosmological constant ($\Lambda$), for a brief review see \cite{Peebles,muhammadsami}; modified gravity, for a good discussion see \cite{Faraoni,Nojiri}; scalar field models (e.g. quintessence, K-essence, tachyon field, phantom (ghost) field \cite{trodden,0264-9381-19-17-311,PhysRevD.70.107301}, dilatonic dark energy, Chaplygin gas) \cite{muhammadsami} and vector fields\footnote{In fact, there is no vector field model to explain the dark energy in the literature but A. Golovnev, V. Mukhanov and V. Vanchurin have described a vector field model for inflation in \cite{vectorinflation} and the same procedure can be used for dark energy too.} \cite{vectorinflation}.

Scalar, vector and tensorial fields came into existence with the development of quantum field theory and were first used by Alan Guth \cite{guth}, Andre Linde \cite{Linde} (plus references therein for discussion), to propose inflation as a possible solution to the horizon and flatness problems. For a review on inflation see \cite{Bassett}.

The homogeneous, isotropic Universe model is described by the Friedmann-Robertson-Walker (FRW) metric and its dynamics is described by the Friedmann equations which are
\begin{eqnarray}
H^2 &=& \dfrac{1}{3}\rho-\dfrac{\kappa}{a^2}~, \label{1stFriedmannequation}
\\
\dfrac{\ddot{a}}{a} &=& -\dfrac{1}{6}\left(1+3\omega_{\text{eff}}\right)\rho~, \label{2ndFriemannequation}
\end{eqnarray}
where $a$ is the scale factor, $H=\dot{a}/a$ is the Hubble parameter, $\rho$ is the total energy density, $\kappa$ is spatial curvature and $P$ is the pressure. Where, $\omega_{\text{eff}}=\Omega_{DE}\text{\space}\omega_{DE}+\Omega_{R}\text{\space}\omega_{R}+\Omega_{M}\text{\space}\omega_{M}$ and the barotropic equation of state $P=\omega\rho$ has been used.  In the equations, natural units $\hbar=c=1$ and $(8\pi G)^{-1/2}=M_P=1$ have been in above and subsequent equations. From eq. (\ref{2ndFriemannequation}), we see that $\ddot{a}>0$ when $\omega_{\text{eff}}<-\frac{1}{3}$.

Using the law of conservation of energy, it can easily be derived that
\begin{eqnarray}\label{conservationofenergy}
\rho\propto a^{-3(1+\omega)}~.
\end{eqnarray}
From eq. (\ref{conservationofenergy}) scalar fields are categorized into two categories on the basis of their equation of state (EoS) parameter $\omega$:
\begin{description}
	\item [The Quintessence field] These fields lie in the domain $-1<\omega<-1/3$. The main feature of this field is that its energy density decrease as in relativistic $(\omega=1/3)$ or non-relativistic matter $(\omega=0)$, but the field could be a possible candidate for dark energy since $\omega<-1/3$ \cite{muhammadsami}.
	\item [The Phantom field] These fields lie in the domain $\omega<-1$ and have the property that their energy density increases as the Universe expands and the scale factor `$a$' becomes infinite in a finite time. These type of fields have negative kinetic energy, which allows $\omega$ to take value less than -1 \cite{muhammadsami}.
\end{description}
When $\omega=-1$ the energy density does not change as the Universe expands and remains constant. The deviation of $\omega$ from $-1$ tells how us energy density, $\rho$, of the field changes as the Universe expands.

The scalar fields used to explain accelerated expansion of the Universe are taken to be new fields (with no connection with Particle Physics). However, the introduction of new fields with no experimental basis other than ``explaining'' one observed phenomenon is too reminiscent of epicycles or the per-relativistic aether. Only if one can exclude fields contained in the standard model, or a minor extension of it, would it be justifiable to introduce such exotic proposals as phantom or quintessence fields, which I argue is necessary and essential to made a unified model of matter and energy.

In this paper, I assume that the present state of the Universe is described by an inert Higgs doublet model. 
If the phase transition in the second SU(2) doublet, $\phi_2$, (which has zero vacuum expectation value (vev)) has not occurred yet, then the component fields of the doublet can be considered as dark energy candidates. 
In the case when the present Universe is described by an inert Higgs doublet model after the second phase transitions different forces and particles would arise (depending upon the VeV of the second doublet) which can be avoided by taking the parameter $m_{22}^2$ of Higgs potential zero thus second phase transition never occurs. The results obtained are presented which favor that accelerated expansion of the Universe is quite possible (in quintessence regime)\footnote{This does not mean that it is the only possible unified model or this has to be the case but certainly it makes its place in the list of dark energy candidacy which has to be checked experimentally.}. Previously, E. Greenwood et al in \cite{PhysRevD.79.103003} considered the appearance of a second vacuum in the SM Higgs potential. They showed that if the second order electroweak phase transition is followed by a first order phase transition (which they were able to achieve for a certain range of parameters of their model) then this phase transition can cause the late time acceleration of the Universe. Carroll et al have also given an independent phantom field model in \cite{trodden} where they have also discussed the stability of energy density along with phantom field as a possible dark energy candidate.
V. K. Onemli and R. P. Woodard have also showed in \cite{0264-9381-19-17-311,PhysRevD.70.107301} that Quantum effects in Cosmology could cause the violation of the null and weak energy conditions (i.e. $\rho+P<0$ in their model) causing $\omega<-1$ on cosmological scales without introducing any ghost, phantom, etc..

The paper is organized as follows; in section 2 we review the inert doublet model and obtain constraints on the inert doublet model's potential to be positive along with the calculations of the minima of the potential. In the next section phase transitions of inert doublet model has been discussed. Section 4 is devoted to obtaining the answer to the question; whether Higgs field can be a possible candidate for dark energy in inert doublet model? In section 5, we discuss the decays of Higgs field(s). Section 6 is where we conclude the paper with future work possibilities.
\section{The inert doublet model}
The electroweak symmetry in the standard model (SM) of Particle Physics is broken spontaneously by the non-zero vev of the Higgs field(s) via Higgs mechanism. The Lagrangian which describes any model in Particle Physics is
\begin{equation}\begin{array}{rcl}\label{L}
\mathscr{L}=\mathscr{L}^{SM}_{gf}+\mathscr{L}_{Y}+\mathscr{L}_{Higgs}~.
\end{array}
\end{equation}
Here, $\mathscr{L}^{SM}_{gf}$ is the $SU_{C}(3){\otimes}SU_{L}(2){\otimes}U_{Y}(1)$ is the SM interaction of the fermions and gauge bosons (force carriers)
\footnote{\begin{eqnarray}\begin{array}{rcl}\label{Lsm}
	\mathscr{L}^{SM}_{gf} &=& -\frac{1}{4}G_{\mu \nu}G^{\mu \nu}-\frac{1}{4}W_{\mu \nu}W^{\mu \nu}-\frac{1}{4}B_{\mu \nu}B^{\mu \nu}
	\\ & &
	+{\bar{\psi}}_{L}^{i}\dot{\iota}{\gamma^{\mu}}{{\nabla}_{\mu}^{EW}}{{\psi}_{L}^{i}}+{\bar{\psi}}_{R}^{i}{\iota}{\sigma^{\mu}}{{\nabla}_{\mu}^{EW}}{{\psi}_{R}^{i}}
	\\ & &
	+{\bar{\chi}}_{L}^{i}{\iota}{\gamma^{\mu}}{{\nabla}_{\mu}^{SM}}{{\chi}_{L}^{i}}+{\bar{U}}_{R}^{i}{\iota}{\sigma^{\mu}}{{\nabla}_{\mu}^{SM}}{{U}_{R}^{i}}+{\bar{D}}_{R}^{i}{\iota}{\sigma^{\mu}}{{\nabla}_{\mu}^{SM}}{{D}_{R}^{i}}~.
	\end{array}
	\end{eqnarray}}
, $\mathscr{L}_{Y}$ is the Yukawa interaction of fermions with the Higgs field(s)\footnote{\vspace{-0.5cm}
	\begin{equation}\begin{array}{rcl}\label{Lyukawa}
	\mathscr{L}_{Y} &=& -Y_{ij}^u{\bar{\chi}}_L^i\tilde{\phi_1}U_R^j-Y_{ij}^d{\bar{\chi}}_L^i{\phi_1}D_R^j-Y_{ij}^e{\bar{\psi}}_L^i{\phi_1}\psi_R^j-h.c.
	\end{array}
	\end{equation}
	here $\psi_L^i$ are left handed leptons doublets, $\psi_R^i$ are right handed leptons singlets, $\chi_L^i$ are left handed quark doublets, $U_R^i$ and $D_R^i$ are the right handed quark singlets. $i$ runs from 1-3. $\phi_1$ is the SM like Higgs doublet.
}
and $\mathscr{L}_{Higgs}$ is the Higgs field Lagrangian where
\begin{equation}\label{LH}
\mathscr{L}_{Higgs}=T_{H}-V_{H}~,
\end{equation}
$T_{H}$ being the kinetic term of the Higgs field(s) and $V_{H}$ being potential of the Higgs field(s)\footnote{
	In the SM one scalar isodoublet with hypercharge $Y=1$ is sufficient to make the theory complete and gauge invariant and hence in SM
	\begin{equation}\begin{array}{rcl}\label{LSMHiggs}
	\mathscr{L}_{Higgs}=T_{H}-V_{H}=(D_{\mu}\phi)^{\dagger}(D^{\mu}\phi)-(-\frac{{\mu}^{2}}{2!}{\phi}^2+\frac{\lambda}{4!}{\phi}^4)
	\end{array}
	\end{equation}
	$$\phi=\begin{pmatrix}
	\phi^+ \\ \phi^0
	\end{pmatrix} \text{\qquad\qquad and\qquad\qquad} \tilde{\phi}=\dot{\iota}\sigma^2\bar{\phi}=\begin{pmatrix}
	\bar{\phi}^0 \\ -\bar{\phi}^+
	\end{pmatrix}.$$}.
Here,
\begin{align}\begin{split}\label{TH}
T_H=({D_1}_{\mu}\phi_1)^\dagger ({D_1}^{\mu}\phi_1) &+ ({D_2}_{\mu}\phi_2)^\dagger ({D_2}^{\mu}\phi_2) \\ &+\left [ {\chi}({D_1}_{\mu}{\phi}_{1})^{\dagger}({D_2}^{\mu}{\phi}_{2})+{\chi^{*}}({D_2}_{\mu}{\phi}_{2})^{\dagger}({D_1}^{\mu}{\phi}_{1}) \right ],
\end{split}\end{align}
and
\begin{align}
V_H &=\text{\space} V_1+V_2+V_{int} ~, \label{V12int}\\
\begin{split}\label{VH}
&=\text{\space}\rho_1\exp(\Lambda_1 m_{11}^2\phi_1^\dagger\phi_1)+\rho_2\exp(\Lambda_2 m_{22}^2\phi_2^\dagger\phi_2)+\rho_3\exp{\Big(}\dfrac{1}{2}\Lambda_3\lambda_1(\phi_1^\dagger\phi_1)^2{\Big)} \\ &\text{\space\space\space} +\rho_4\exp{\Big(}\dfrac{1}{2}\Lambda_4\lambda_2(\phi_2^\dagger\phi_2)^2{\Big)}+{m_{12}^2}(\phi_1^\dagger \phi_2)+{{m_{12}^2}^*}(\phi_2^\dagger \phi_1)
+\lambda_3 (\phi_1^\dagger\phi_1)(\phi_2^\dagger\phi_2)
\\ &\text{\space\space\space}
+\lambda_4(\phi_1^\dagger\phi_2)(\phi_2^\dagger\phi_1)+\dfrac{1}{2}{\big[}\lambda_5(\phi_1^\dagger\phi_2)^2 +\lambda_5^* (\phi_2^\dagger\phi_1)^2{\big]}
+\lambda_6(\phi_1^\dagger\phi_1)(\phi_1^\dagger\phi_2)
\\ &\text{\space\space\space}
+\lambda_6^*(\phi_1^\dagger\phi_1)(\phi_2^\dagger\phi_1)+\lambda_7(\phi_2^\dagger\phi_2)(\phi_1^\dagger\phi_2)+\lambda_7^*(\phi_2^\dagger\phi_2)(\phi_2^\dagger\phi_1)~,
\end{split}
\end{align}
where
$V_1$ and $V_2$ in eq. (\ref{V12int}) are the Lagrangian of Higgs field $\phi_1$ (given by first two terms of RHS of eq. (\ref{VH})) and $\phi_2$ (given by $3^{rd}$ and $4^{th}$ terms of RHS of eq. (\ref{VH})) respectively, $V_{int}$ is the interaction Lagrangian of fields $\phi_1$ and $\phi_2$, given by remaining terms,
$${D_1}_\mu=\partial_\mu+\dot{\iota}\dfrac{g_1}{2}\sigma_i {W^i}_\mu+\dot{\iota}\dfrac{g_1'}{2}B_\mu ~,$$
$${D_2}_\mu=\partial_\mu+\dot{\iota}\dfrac{g_2}{2}\sigma_i {W^i}_\mu+\dot{\iota}\dfrac{g_2'}{2}B_\mu ~,$$
$$
\phi_{i}=
\begin{bmatrix}
\phi^{+}_{i} \\
\eta_i + \dot{\iota}\chi_i +\nu_i \\
\end{bmatrix},
\text{\qquad \quad \qquad}
\phi_{i}^{\dagger}=
\begin{bmatrix}
\phi^{-}_{i} & \eta_i - \dot{\iota}\chi_i +\nu_i
\end{bmatrix}.
$$
The dimensions of different parameters are
$$[\rho_i]^{-1}=[\Lambda_i]=[L]^4,\text{\space} [m_{ii}^2]=[L]^{-2},\text{\space} [\phi_i]=[L]^{-1}\text{\space and \space}[\lambda_i]=[L]^0.$$
Here ``$L$'' denotes the length. The fields $\phi^{+}_{i}$, $\phi^{-}_{i}$, $\eta_i$ and $\chi_i$ are the hermitian Higgs fields ($\phi^{\pm}_{i}$ are charged whereas other fields are neutral), $\nu_i$ is the vev of the doublet $\phi_i$.

The Higgs vacuum energy of the potential given by eq. (\ref{VH}) is 
\begin{equation}\label{Evac}
\begin{array}{rcl}
E_{vac}&=&\text{\space\space}\rho_1 \exp\left(\dfrac{1}{2}m_{11}^2\Lambda_1\nu_1^2\right)+\rho_2 \exp\left(\dfrac{1}{2}m_{22}^2 \Lambda_2\nu_2^2\right)+\rho_3 \exp\left(\dfrac{1}{8}\lambda_1\Lambda_3\nu_1^4\right) \\ && +\rho_4 \exp\left(\dfrac{1}{8}\lambda_2\Lambda_4\nu_2^4\right)+\dfrac{1}{4}\lambda_3\nu_1^2\nu_2^2+\dfrac{1}{4}\lambda_4\nu_1^2\nu_2^2+\dfrac{1}{8}\left(\lambda_5+\lambda_5^*\right)\nu_1^2\nu_2^2 \\ && +\dfrac{1}{4}\left(\lambda _6+\lambda _6^*\right)\nu _1^3\nu _2+ \dfrac{1}{4} \left(\lambda_7+\lambda_7^*\right) \nu_1 \nu_2^3.
\end{array}
\end{equation}
We would be choosing $\nu_2=0$ here (explained in section \ref{MtHp}), which corresponds to the inert doublet model, then
\begin{equation}\label{Evacinertmodel}
\begin{array}{rcl}
E^{'}_{vac}&=&\rho_2+\rho_4+\rho_1 \exp\left(\dfrac{1}{2}m_{11}^2\Lambda_1\nu_1^2\right)+\rho_3 \exp\left(\dfrac{1}{8}\lambda_1\Lambda_3\nu_1^4\right).
\end{array}
\end{equation}	

Since we are looking to provide the accelerated expansion to the Universe from Higgs field(s) we want the Higgs field(s) to live longer than the age of the Universe (or at least not less than the present age of the Universe). Thus, stable Higgs field(s) is(are) required for the scalar field dark energy model. This can be achieved by imposing a discrete $Z_2$ symmetry $\phi\rightarrow-\phi$. When $Z_2$ symmetry is broken by $(\phi_i^\dagger\phi_j)$ type terms then it is said to be softly broken and when $Z_2$ symmetry is broken by $(\phi_i^\dagger\phi_j)(\phi_k^\dagger\phi_l)$ type terms then symmetry is said to be hardly broken.
The terms containing $m_{12}^{2}$ 
describe the soft symmetry breaking of $Z_{2}$ symmetry, the terms $\lambda_{6}$ and $\lambda_{7}$ describe the hard symmetry breaking of $Z_2$ symmetry. In the absence of these terms along with no cross kinetic term i.e. $\chi=0$, the 2HDM's Higgs Lagrangian has a perfect $Z_{2}$ symmetry \cite{Ginzburg-2}. There are two $Z_2$ symmetries corresponding to the doublets:
\begin{eqnarray}
\text{I:\qquad\qquad}\phi_1 \longrightarrow &-\phi_1~, \text{\qquad\qquad} \phi_2 \longrightarrow &\text{\space\space}\phi_2~, \label{1stZ2symmetry}
\\
\text{II:\qquad\qquad}\phi_1 \longrightarrow &\text{\space\space}\phi_1~, \text{\qquad\qquad} \phi_2 \longrightarrow &-\phi_2~.	\label{2ndZ2symmetry}
\end{eqnarray}
For $Z_{2}$ symmetry see \cite{Ginzburg-2}.

It should also be noted that even if the Higgs field(s) has(have) decayed off 
its vacuum energy given by eq. (\ref{Evac}) would still be able to provide accelerated expansion provided it dominates at some time in the life span of the Universe.
\subsection{Positivity constraint on Higgs potential}
The stability of the vacuum given by eq. (\ref{VH}) implies that the potential must be positive for all asymptotically large values of the Higgs field. Even though $V_1$ and $V_2$ remains positive for all values of fields considering $\rho_i>0$, we should find out the constraints on the parametric product $\Lambda_1 m_{11}^2$, $\Lambda_2 m_{22}^2$, $\frac{1}{2}\Lambda_3 \lambda_1$ and $\frac{1}{2}\Lambda_4 \lambda_2$ when the exponentials are approximated upto a finite order ($4$th here). The conditions obtained should match on to the conditions for the general 2HDM potential given in \cite{Ginzburg-2}.

Some conditions can be obtained to achieve the stability of the vacuum by introducing new parameters as \cite{2hdmc,kaffaskhater},
$$\left|\phi_1\right|=r \cos\gamma~,\text{\qquad}\left|\phi_2\right|=r \sin\gamma~,\text{\qquad}\dfrac{\phi_2^\dagger\phi_1}{\left|\phi_1\right|\left|\phi_2\right|}=\rho \exp(\dot{\iota}\theta)~,$$
where $\gamma\in[0,\pi / 2]$, $\rho\in [0,2\pi)$. Using these transformations in the potential within its expanded form upto forth order (with $\rho_i>0$ and hence will be omitted in eq. (\ref{positivityconstraintpotential})), the potential becomes
\begin{equation}
\begin{array}{rcl}\label{positivityconstraintpotential}
V &=& r^4\Big( \lambda_1^{'}\cos^4\gamma+\lambda_2^{'}\sin^4\gamma+\lambda_3\cos^2\gamma\sin^2\gamma+\lambda_4\rho^2\cos^2\gamma\sin^2\gamma \\ &&
~~~~+\lambda_5\rho^2\cos^2\gamma\sin^2\gamma\cos2\theta+2\rho\cos\gamma\sin\gamma\cos\theta[\lambda_6\cos^2\gamma+\lambda_7\sin^2\gamma] ~ \Big)~,
\end{array}
\end{equation}
the positivity of eq. (\ref{positivityconstraintpotential}) implies that
\begin{eqnarray}\label{positivitycontraint1}
\lambda_1^{'}>0~, \text{\qquad} \lambda_2^{'}>0~, \text{\qquad}\lambda_3>-2~\sqrt{\lambda_1^{'}\lambda_2^{'}}~,
\end{eqnarray}
when $\lambda_6=\lambda_7=0$, then we also have
\begin{eqnarray}\label{positivitycontraint2}
\lambda_3+\text{min}[0,\lambda_4-\left| \lambda_5\right|] >-2~\sqrt{\lambda_1^{'}\lambda_2^{'}}~,
\end{eqnarray}
where $\lambda_1^{'}=\dfrac{1}{2}\left(\rho_3\Lambda_3\lambda_1+\rho_1(\Lambda_1 m_{11}^2)^2\right)$ and $\lambda_2^{'}=\dfrac{1}{2}\left(\rho_4\Lambda_4\lambda_2+\rho_2(\Lambda_2 m_{22}^2)^2\right)$.
\subsection{Minimizing the Higgs potential}\label{MtHp}
The extrema of the potential are found by taking
\begin{equation}\label{extremaconditions}
\dfrac{\partial V_H}{\partial \phi_1} {\bigg|}_{\substack{
		\phi_1=\left\langle \phi_1\right\rangle \\
		\phi_2=\left\langle \phi_2\right\rangle
	}}=\dfrac{\partial V_H}{\partial \phi_1^\dagger}\bigg |_{\substack{
	\phi_1=\left\langle \phi_1\right\rangle \\
	\phi_2=\left\langle \phi_2\right\rangle
}}=0 \text{\qquad and \qquad} \dfrac{\partial V_H}{\partial \phi_2}\bigg |_{\substack{
\phi_1=\left\langle \phi_1\right\rangle \\
\phi_2=\left\langle \phi_2\right\rangle
}}=\dfrac{\partial V_H}{\partial \phi_2^\dagger}\bigg |_{\substack{
\phi_1=\left\langle \phi_1\right\rangle \\
\phi_2=\left\langle \phi_2\right\rangle
}}=0~.
\end{equation}		
The most general solution of the conditions (\ref{extremaconditions}) is
\begin{equation*}\label{VeV}
\left\langle \phi_1\right\rangle =\frac{1}{\sqrt{2}}\begin{pmatrix}
0 \\ \nu_1
\end{pmatrix}
\text{\qquad and \qquad}
\left\langle \phi_2\right\rangle =\frac{1}{\sqrt{2}}\begin{pmatrix}
u \\ \nu_2
\end{pmatrix}.
\end{equation*}
The first solution of extrema has been taken to be similar to the Higgs vacuum in SM and the second one is the most general which can occur. One needs to keep in mind that now $\nu^2=\nu_1^2+\left| \nu_2^2\right|+u^2$. In the electroweak symmetry breaking situation where $\left\langle \phi_1\right\rangle \neq 0 $, choosing the z-axis in weak isospin space such that $\nu_1\ge 0$. 

When $u\neq 0$, the non-zero value of $u$ will contribute to the ``charged'' type dark energy, which has not been observed. To avoid this, we would take $u=0$. From the extrema conditions given by eq. (\ref{extremaconditions}), we can determine the values of $\nu_1$ and $\nu_2$ \cite{2hdmc,Ginzburg-1}, solving eq. (\ref{extremaconditions}) for the potential given by eq. (\ref{VH}) leads to
\begin{equation*}\begin{array}{rcl}\label{1stconditionwithexponential}
\nu_1 {\Big[}2 \rho_1 \Lambda_1 m_{11}^2\exp{\Big(}\dfrac{\Lambda_1 m_{11}^2}{2}\nu_1^2{\Big)} &+& \rho_3\Lambda_3\lambda_1\nu_1^2\exp{\Big(}\dfrac{\Lambda_3\lambda_1}{8}\nu_1^4{\Big)}-2\text{Re}(m_{12}^2)\frac{\nu_2}{\nu_1}\\ &+& (\lambda_3+\lambda_4+\text{Re}(\lambda_5))\nu_2^2+3\text{Re}(\lambda_6)\nu_1\nu_2+\text{Re}(\lambda_7)\dfrac{\nu_2^3}{\nu_1}{\Big]}=0~,
\end{array}
\end{equation*}
\begin{equation*}\begin{array}{rcl}\label{2ndconditionwithexponential}
\nu_2{\Big[} 2\rho_2 \Lambda_2 m_{22}^2\exp{\Big(}\dfrac{\Lambda_2 m_{22}^2}{2}\nu_2^2{\Big)} &+& \rho_4\Lambda_4\lambda_2\nu_2^2\exp{\Big(}\dfrac{\Lambda_4\lambda_2}{8}\nu_2^4{\Big)}-2\text{Re}(m_{12}^2)\frac{\nu_1}{\nu_2} \\ &+& (\lambda_3+\lambda_4+\text{Re}(\lambda_5))\nu_1^2
+\text{Re}(\lambda_6)\dfrac{\nu_1^3}{\nu_2}+3\text{Re}(\lambda_7)\nu_1\nu_2{\Big]}=0~.
\end{array}
\end{equation*}
The solution of the above equations for $\nu_1$ and $\nu_2$ is practically impossible since the exponentials contain $\nu_1$ and $\nu_2$. 
The  vev's are approximated by truncating the potential eq. (\ref{VH}) upto forth order. This is also a good approximation in the sense that the results of $\nu's$ would also be true for the general 2HDM potential vev's of \cite{Ginzburg-2}. Applying minimization conditions given by eq. (\ref{extremaconditions}) we get, 
\begin{equation}\begin{array}{rcl}\label{1stcondition}
\nu_1\left[ 2 \rho_1 \Lambda_1 m_{11}^2+2\lambda_1^{'}\nu_1^2-2\text{Re}(m_{12}^2)\dfrac{\nu_2}{\nu_1} \right. &+& (\lambda_3+\lambda_4+\text{Re}(\lambda_5))\nu_2^2
\\ &+& \left. 3\text{Re}(\lambda_6)\nu_1\nu_2+\text{Re}(\lambda_7)\dfrac{\nu_2^3}{\nu_1}\right]=0~,
\end{array}
\end{equation}
\begin{equation}\begin{array}{rcl}\label{2ndcondition}
\nu_2\left[ 2 \rho_2 \Lambda_2 m_{22}^2+2\lambda_2^{'}\nu_2^2-2\text{Re}(m_{12}^2)\dfrac{\nu_1}{\nu_2} \right. &+&(\lambda_3+\lambda_4+\text{Re}(\lambda_5))\nu_1^2
\\ &+& \left. \text{Re}(\lambda_6)\dfrac{\nu_1^3}{\nu_2}+3\text{Re}(\lambda_7)\nu_1\nu_2\right]=0~,
\end{array}
\end{equation}
where $\lambda_1^{'}=\dfrac{1}{2}\left(\rho_3\Lambda_3\lambda_1+\rho_1(\Lambda_1 m_{11}^2)^2\right)$ and $\lambda_2^{'}=\dfrac{1}{2}\left(\rho_4\Lambda_4\lambda_2+\rho_2(\Lambda_2 m_{22}^2)^2\right)$ as given before.

Since $Z_2$ symmetry is very important in determining the phenomenology of the theory and with exact $Z_2$ symmetry the lightest Higgs boson (field) is stable. 
In this article, the potential which is $Z_2$ symmetric will also be taken and the possibility of any Higgs field(s) to be the candidate for the dark energy will be examined, if possible it would give an ever expanding Universe while accelerating. Thus, imposing the $Z_2$ symmetry \cite{Ginzburg-2} we have,
\begin{equation}\label{Z2Symmetry}
\chi=0 \text{ ,\qquad} m_{12}^2=0 \text{ ,\qquad} \lambda_6=0 \text{ ,\qquad} \lambda_7=0~,
\end{equation}
Using eqs. (\ref{1stcondition}, \ref{2ndcondition}, \ref{Z2Symmetry}), we get four solutions for $\nu_1$ and $\nu_2$, which are
\begin{eqnarray}
\nu_1^2 &= 0 ~, \text{\qquad\qquad\qquad\qquad\qquad\qquad\qquad} \nu_2^2 &= 0 ~, \label{electroweaksymmetricvacuum}
\\
\nu_1^2 &= 0 ~, \text{\qquad\qquad\qquad\qquad\qquad\qquad\qquad} \nu_2^2 &= -\dfrac{\rho_2 \Lambda_2 m_{22}^2}{\lambda_2^{'}} ~, \label{inertvacuum1}
\\
\nu_1^2 &= -\dfrac{\rho_1 \Lambda_1 m_{11}^2}{\lambda_1^{'}} ~, \text{\qquad\qquad\qquad\qquad\quad\space\space\space} \nu_2^2 &= 0 ~, \label{inertvacuum2}
\\
\nu_1^2 &= -\dfrac{2(2\rho_1 \Lambda_1\lambda_2^{'} m_{11}^2-\rho_2 \Lambda_2 \lambda_{345}m_{22}^2)}{4\lambda_1^{'} \lambda_2^{'}-\lambda_{345}^2} ~, \text{\quad} \nu_2^2 &= -\dfrac{2(2\rho_2 \Lambda_2 \lambda_1^{'} m_{22}^2-\rho_1 \Lambda_1\lambda_{345}m_{11}^2)}{4\lambda_1^{'} \lambda_2^{'}-\lambda_{345}^2} ~, \label{mixedvacuum}
\end{eqnarray}
where $\lambda_{345}=\lambda_3+\lambda_4+\text{Re}(\lambda_5)$.
\paragraph*{Electroweak symmetric vacuum} This vacuum solution is given by eq. (\ref{electroweaksymmetricvacuum}) represents the electroweak symmetric extremum, in this case the complete Lagrangian respects $Z_2$ symmetry with respect to both fields given by eq. (\ref{1stZ2symmetry}) and eq. (\ref{2ndZ2symmetry}) i.e. I and II, here $m_{11}^2>0$ and $m_{22}^2>0$. The doublets are
$$
\phi_{i}=
\begin{bmatrix}
\phi^{+}_{i} \\
\eta_i + \dot{\iota}\chi_i \\
\end{bmatrix}
\text{\qquad and \qquad}
\phi_{i}^{\dagger}=
\begin{bmatrix}
\phi^{-}_{i} & \eta_i - \dot{\iota}\chi_i
\end{bmatrix}.
$$
\paragraph*{Inert vacuum-I}
Eq. (\ref{inertvacuum1}) will be called inert vacuum-I. The doublets in this vacuum are
$$
\phi_{1}=
\begin{bmatrix}
\phi^{+}_{1} \\
\eta_1 + \dot{\iota}\chi_1 \\
\end{bmatrix}
\text{\qquad and \qquad}
\phi_{2}=
\begin{bmatrix}
\phi^{+}_{2} \\
\eta_2 + \dot{\iota}\chi_2 + \nu_2 \\
\end{bmatrix},
$$
in this type of vacuum solution, the fields $\phi_2^\pm$ and $\chi_2$ are the Goldstone bosons and all other fields are physical. In this case, the complete Lagrangian respects $Z_2$ symmetry given by eq. (\ref{1stZ2symmetry}) and only the Higgs Lagrangian (not the complete Lagrangian because of the Yukawa Lagrangian since $\phi_2$ is not contained in Yukawa Lagrangian) violates given by eq. (\ref{2ndZ2symmetry}).
\paragraph*{Inert vacuum-II}
The doublets in this type of vacuum, given by eq. (\ref{inertvacuum2}), are
\begin{equation}
\phi_{1}=
\begin{bmatrix}
\phi^{+}_{1} \\
\eta_1 + \dot{\iota}\chi_1 + \nu_1 \\
\end{bmatrix},
\text{\qquad\qquad}
\phi_{2}=
\begin{bmatrix}
\phi^{+}_{2} \\
\eta_2 + \dot{\iota}\chi_2\\
\end{bmatrix},
\end{equation}
the fields $\phi_1^\pm$ and $\chi_1$ are the Goldstone bosons whereas other fields are physical. The Yukawa interaction is described by the interaction of field $\phi_1$ with fermions ($\phi_2$ does not couple to fermions but it can appear in loop process).

The Higgs and Yukawa Lagrangian violates $Z_2$ symmetry given by eq. (\ref{1stZ2symmetry}) whereas respects given by eq. (\ref{2ndZ2symmetry}). Thus, parity in field $\phi_2$ is conserved this makes the lightest particle/field of doublet $2$ stable. If there is no phase transitions (which brings Universe in the mixed vacuum state) in the future of the Universe or before the phase transitions in doublet $\phi_2$ then this lightest particle is a good candidate for dark energy, giving rise to an ever accelerating expanding Universe. 
Without the stability, we would have a model for accelerated expanding Universe for a limited time depending upon the decay width (life time) of the dark energy field(s), because if field(s) can decay then it ceases to provide accelerated expansion after its decay to the other fields only if its vacuum energy has not dominates before its decay.
\\
The masses of the fields in this vacuum are
\begin{equation}\begin{array}{rcl}\label{inertvacuumHiggsmasses}
&&
m_{\eta_1}^2=2\lambda_1^{'}\nu_1^2~,
\\ &&
m_{\eta_2}^2=\rho_2\Lambda_2m_{22}^2+\dfrac{\lambda_3+\lambda_4+\text{Re}(\lambda_5)}{2}\nu_1^2~,
\\ &&
m_{\chi_2}^2=\rho_2\Lambda_2m_{22}^2+\dfrac{\lambda_3+\lambda_4-\text{Re}(\lambda_5)}{2}\nu_1^2~,
\\ &&
m_{\phi_2^{\pm}}^2=\rho_2\Lambda_2m_{22}^2+\dfrac{\lambda_3}{2}\nu_1^2~,
\end{array}
\end{equation}
where $\nu_1=\dfrac{1}{\sqrt[4]{2{G_F}^2}}\approx246$GeV.
\\
In this paper, I am interested in this type of vacuum. 
\paragraph*{Mixed vacuum} After symmetry is broken, the Lagrangian violates the $Z_2$ symmetry given by eqs. (\ref{1stZ2symmetry}) and (\ref{2ndZ2symmetry}).
The physical fields in this model are the combination of fields from $\phi_1$ and $\phi_2$. The complete theory and phenomenology of this model is discussed in \cite{Brancoetal}. Mixed vacuum solution is given by eq. (\ref{mixedvacuum}).
\section{Phase transitions in inert doublet model/two Higgs doublet model}
When we look for dark energy to be some physical field(s), then it becomes essential to also look for its evolution in the history (cooling) of the Universe.

According to \cite{Ginzburg-1}, in the quantum field theory at non-zero temperature 
the terms $m_{ii}^2$ of the quadratic terms evolve with temperature as
\begin{eqnarray}
&& m_{11}^2\longrightarrow m_{11}^2+\dfrac{1}{2}c_1 T^2, \label{m112thermalevolution} \\ &&
m_{22}^2\longrightarrow m_{22}^2+\dfrac{1}{2}c_2 T^2, \label{m222thermalevolution}
\end{eqnarray}
where
$$c_1=\dfrac{6\lambda_1^{'}+2\lambda_3+\lambda_4}{12}+\dfrac{3g_1^2+g_1'^2}{32}+\dfrac{g_t^2+g_b^2}{8},$$
and
$$c_2=\dfrac{6\lambda_2^{'}+2\lambda_3+\lambda_4}{12}+\dfrac{3g_2^2+g_2'^2}{32},$$
here
$\lambda_1^{'}=\dfrac{1}{2}\left(\rho_3\Lambda_3\lambda_1+\rho_1(\Lambda_1 m_{11}^2)^2\right)$, $\lambda_2^{'}=\dfrac{1}{2}\left(\rho_4\Lambda_4\lambda_2+\rho_2(\Lambda_2 m_{22}^2)^2\right)$ as above and 
$g_{1,2}$ and $g'_{1,2}$ are the Electroweak gauge couplings with doublet $\phi_1$ and $\phi_2$ respectively, $g_t\approx 1$ and $g_b\approx 0.03$ are the top and bottom quark Yukawa couplings with the doublet $\phi_1$ in inert doublet model respectively.
\\
In general, $c_1$ and $c_2$ can have any sign but the potential positivity implies that (in any situation)
\begin{equation*}
c_1+c_2>0,
\end{equation*}
with the above mentioned positivity constraints, the case when $c_1>0$ and $c_2>0$ will be considered.
\\
The Higgs potential with the new quadratic terms now become
\begin{equation}\begin{array}{rcl}\label{VHT}
V_H(\phi_1,\phi_2,T)&=&E^{'}_{vac}+\dfrac{1}{2}c_1(T^2-{T_c}_1^2)(\phi_1^\dagger\phi_1)+\dfrac{1}{2}c_2(T^2-{T_c}_2^2)(\phi_2^\dagger\phi_2)
\\ & &
+\lambda_1^{'}(\phi_1^\dagger\phi_1)^2+\lambda_2^{'}(\phi_2^\dagger\phi_2)^2+{m_{12}^2}(\phi_1^\dagger\phi_2)+{m_{12}^2}^{*}(\phi_2^\dagger\phi_1)
\\ & &
+\lambda_3 (\phi_1^\dagger\phi_1)(\phi_2^\dagger\phi_2)+\lambda_4(\phi_1^\dagger\phi_2)(\phi_2^\dagger\phi_1)+\dfrac{1}{2}[\lambda_5(\phi_1^\dagger\phi_2)^2 +\lambda_5^* (\phi_2^\dagger\phi_1)^2]
\\ & &
+(\phi_1^\dagger\phi_1)[ \lambda_6(\phi_1^\dagger\phi_2)+\lambda_6^*(\phi_2^\dagger\phi_1) ]+(\phi_2^\dagger\phi_2)[ \lambda_7(\phi_1^\dagger\phi_2)+\lambda_7^*(\phi_2^\dagger\phi_1) ]
\\ & &
+\text{higher order terms}~,
\end{array}
\end{equation}
In this case corresponding to the two different vev's of the fields $\phi_1$ and $\phi_2$, there would be occurring two different critical temperatures $T_{c_{1}}=\sqrt{-\dfrac{2~\rho_1 \Lambda_1 m_{11}^2}{c_1}}$ and $T_{c_{2}}=\sqrt{-\dfrac{2~\rho_2 \Lambda_2 m_{22}^2}{c_2}}$ respectively. In this analysis we require that the phase transitions in doublet $\phi_2$ occur after phase transitions in $\phi_1$, thus $T_{c_{1}}>T_{c_{2}}$.
\\~\\
There are three possible situations depending upon the value of temperature,
\begin{description}
	\item[$T>T_{c_{1}}$] \hfill \\
	For $T>T_{c_{1}}$ the symmetry is not broken in either field and the phase transitions has not occurred. Thus, in this situation minima of both fields lies at zero given by eq. (\ref{electroweaksymmetricvacuum}). All fields are massless in this temperature range i.e.$T\geq {T_c}_1$. Since, symmetry is not broken in this limit all the fermions and bosons are also massless.
	\item[$T_{c_{2}}<T<T_{c_{1}}$] \hfill \\
	This is the most interesting situation as the symmetry is retained in field $\phi_2$ and is broken in $\phi_1$. In this situation due to phase transitions in $\phi_1$ all fields (fermions and all bosons) gets masses, but since the phase transitions in field $\phi_2$ has not occurred so it still is a possible candidate for dark energy field. 
	This situation is also interesting because this article assumes that the present observed Universe happens to occur in this domain.
	The expansion of $\phi_1$ around its minimum $\nu_1$ leads to exactly the same phenomenology as that of the SM.
	\item[$T<T_{c_{2}}$] \hfill \\
	When $T<T_{c_{2}}$ symmetry has got broken spontaneously in both fields $\phi_1$ and $\phi_2$ at temperatures ${T_c}_1$ and ${T_c}_2$ respectively. In this situation, both fields gets perturbed around their true minima. This perturbation give masses to all the mixed fields of the doublets $\phi_1$ and $\phi_2$ (except the CP odd and charged field of angle $\beta=\tan^{-1}\frac{\nu_2}{\nu_1}$ rotated new Higgs doublet $H_1$ which are the Goldstone bosons). 
	This should be expected to occur some time in the future considering $m_{22}^2\neq 0$.		
\end{description}
\subsection*{Constraint on parameters from phase transitions}
If we require that the phase transitions in the doublet $\phi_1$ (which is acting now to be SM Higgs doublet) occur at the same temperature as in SM Higgs doublet, we can constrain the parameters of the potential. Note that this condition is not necessary but an interesting one which can probe this model (one can also relax this condition). The critical temperature at which phase transitions occur in SM is \cite{kapusta},
$$T_c^2=\dfrac{4~\lambda~\nu^2}{2\lambda+\dfrac{3}{4}g^2+\dfrac{1}{4}{g^{'}}^2}$$
\\
where $\lambda=0.1305$ is the quartic coupling of Higgs doublet in SM, $g=0.6376$ is the Higgs doublet coupling with SU(2) gauge group and $g^{'}=0.3441$ is the Higgs doublet coupling with U(1) gauge group. With the values of parameters given above we get $T_c=230.3186$GeV.
\\
Setting $T_{c_1}=T_c$ gives,
\begin{equation*}
\sqrt{\dfrac{-2~\rho_1~\Lambda_1~m_{11}^2}{\dfrac{6\lambda_1^{'}+2\lambda_3+\lambda_4}{12}+\dfrac{3g_1^2+g_1'^2}{32}+\dfrac{g_t^2+g_b^2}{8}}}=\sqrt{\dfrac{4~\lambda~\nu^2}{2\lambda+\dfrac{3}{4}g^2+\dfrac{1}{4}{g^{'}}^2}}~,
\end{equation*}
taking square and using equation (\ref{inertvacuum2}) in,
\begin{equation*}
\dfrac{2~\lambda_1^{'}~\nu_1^2}{\dfrac{6\lambda_1^{'}+2\lambda_3+\lambda_4}{12}+\dfrac{3g_1^2+g_1'^2}{32}+\dfrac{g_t^2+g_b^2}{8}}=\dfrac{4~\lambda~\nu^2}{2\lambda+\dfrac{3}{4}g^2+\dfrac{1}{4}{g^{'}}^2}~,
\end{equation*}	
here $\nu^2=\nu_1^2={246~\text{GeV}}^2$, $\lambda_i$'s are the parameters of inert doublet model potential, $g_1$ and $g_1'$ are the couplings of doublet $\phi_1$ with the SU(2) and U(1) gauge group respectively. Simplification of the above equation with the numerical values lead to
\begin{equation}\label{phasetransitionsconstraint}
2\lambda_3+\lambda_4=0.7869~,
\end{equation}
Here, in deriving the eq. (\ref{phasetransitionsconstraint}), $\lambda_1^{'}=0.1305$, $g_1=0.6376$, $g_1'=0.3441$, $g_t=1$ and $g_b=0.03$ have been used.
\section{Higgs field(s) as dark energy field(s)}
The second Friedmann eq. (\ref{2ndFriemannequation}) tells that accelerated expansion will occur when $\omega_{\text{eff}}<-\frac{1}{3}$. For the field $\phi_2$ to be the dark energy field, it must bring $\omega_{\text{eff}}<-\frac{1}{3}$ in the history of Universe (in fact just now $Z\approx 0.37$ with $\Omega_\text{m}=0.3$ and $\Omega_{\text{DE}}=0.7$) which tells that the symmetry in this field should not be broken. For this purpose, we need to solve the Euler Lagrange equations in some background, which are
\begin{eqnarray}\label{EulerLagrangeEquations}
\partial_\mu\left( \dfrac{\partial(\sqrt{-g}{\mathscr{L}_{Higgs})}}{\partial(\partial_\mu \psi_i)} \right)-\dfrac{\partial(\sqrt{-g}{\mathscr{L}_{Higgs}})}{\partial \psi_i}=0,
\end{eqnarray}
where $\psi_i$ are different fields of doublets $\phi_1$ and $\phi_2$.
\\
The solution of Euler Lagrange equations of motion in FRW Universe $(\sqrt{-g}=a(t)^3)$ for the fields $\phi_{2}^\pm$, $\eta_2$ and $\chi_2$ given above is
\begin{equation}\begin{array}{rcl}\label{etaequationofmotion}
& & \ddot{\eta}_2+3\dfrac{\dot{a}}{a}\dot{\eta}_2+\dfrac{1}{2}\left[\eta_2\left(\left(\lambda_3+\lambda_4+\lambda_5\right)\nu^2+2m_{22}^2\Lambda_2\rho_2\exp\left(\dfrac{1}{2}m_{22}^2\Lambda_2\left(\chi_2^2+\eta_2^2+2{\phi^{c}_{2}}^2\right)\right)\right.\right. \\
& & \left.+\lambda_2\Lambda_4\rho_4\chi_2^2\exp\left(\dfrac{1}{8}\lambda_2\Lambda_4\left(\chi_2^2+\eta_2^2+2{\phi^{c}_{2}}^2\right)^2\right)+2\lambda_2\Lambda_4\rho_4{\phi^{c}_{2}}^2\exp\left(\dfrac{1}{8}\lambda_2\Lambda_4\left(\chi_2^2+\eta_2^2+2{\phi^{c}_{2}}^2\right)^2\right)\right) \\
& & \left.+\lambda_2\Lambda_4\rho_4\eta_2^3\exp\left(\dfrac{1}{8}\lambda_2\Lambda_4\left(\chi_2^2+\eta_2^2+2{\phi^{c}_{2}}^2\right)^2\right)\right]=0 \\
\end{array}
\end{equation}	
\begin{equation}\begin{array}{rcl}\label{chiequationofmotion}
& & \ddot{\chi}_2+3\dfrac{\dot{a}}{a}\dot{\chi}_2+\dfrac{1}{2}\left(\chi_2\left(\left(\lambda_3+\lambda_4-\lambda_5\right)\nu^2+2m_{22}^2\Lambda_2\rho_2\exp\left(\dfrac{1}{2}m_{22}^2\Lambda_2\left(\chi_2^2+\eta_2^2+2{\phi^{c}_{2}}^2\right)\right)\right.\right. \\ & & \left.+\lambda_2\Lambda_4\rho_4\eta_2^2\exp\left(\dfrac{1}{8}\lambda_2\Lambda_4\left(\chi_2^2+\eta_2^2+2{\phi^{c}_{2}}^2\right)^2\right)+2\lambda_2\Lambda_4\rho_4{\phi^{c}_{2}}^2\exp\left(\dfrac{1}{8}\lambda_2\Lambda_4\left(\chi_2^2+\eta_2^2+2{\phi^{c}_{2}}^2\right)^2\right)\right) \\ & & \left.+\lambda_2\Lambda_4\rho_4\chi_2^3\exp\left(\dfrac{1}{8}\lambda_2\Lambda_4\left(\chi_2^2+\eta_2^2+2{\phi^c_2}^2\right)^2\right)\right)=0
\end{array}
\end{equation}
\begin{equation}\begin{array}{rcl}\label{chargedHiggsequationofmotion}
& & \ddot{\phi}^{c}_{2}+3\dfrac{a'}{a}\dot{\phi}^{c}_{2}+\dfrac{1}{2}\left(\lambda_3\nu^2+2m_{22}^2\Lambda_2\rho_2\exp\left(\dfrac{1}{2}m_{22}^2\Lambda_2\left(\chi_2^2+\eta_2^2+2{\phi^{c}_{2}}^2\right)\right)\right. \\ & & \left.+\lambda_2\Lambda_4\rho_4\left(\chi_2^2+\eta_2^2+2{\phi^{c}_{2}}^2\right)\exp\left(\dfrac{1}{8}\lambda_2\Lambda_4\left(\chi_2^2+\eta_2^2+2{\phi^{c}_{2}}^2\right)^2\right)\right)\phi^{c}_{2}=0
\end{array}
\end{equation}
\\
To avoid the confusion between positively and negatively charged Higgs boson, `c' instead of $+$ or $-$ has been used. 
\\
The energy density and pressure after expansion of 2HDM Higgs Lagrangian for physical fields is
\begin{equation}\begin{array}{rcl}\label{rhoHiggsPHiggs}
\rho_{Higgs}/P_{Higgs} & = & \dfrac{1}{2}\dot{\eta_2^2}+\dfrac{1}{2}\dot{\chi_2^2}+\dot{\phi_2^c}^2\pm\left[\dfrac{1}{4}\left(\lambda_3+\lambda_4+\lambda_5\right)\nu^2\eta_2^2+\dfrac{1}{4}\left(\lambda_3+\lambda_4-\lambda_5\right)\nu^2\chi_2^2\right. \\ & & +\dfrac{1}{2}\lambda_3\nu^2{\phi_2^c}^2+\rho_2\exp\left(\dfrac{1}{2}m_{22}^2\Lambda_2\chi_2^2+\dfrac{1}{2}m_{22}^2\Lambda_2\eta_2^2+m_{22}^2\Lambda_2{\phi_2^c}^2\right) \\ & & +\rho_4\exp\left(\dfrac{1}{8}\lambda_2\Lambda_4\chi_2^4+\dfrac{1}{8}\lambda_2\Lambda_4\eta_2^4+\dfrac{1}{2}\lambda_2\Lambda_4{\phi_2^c}^4+\dfrac{1}{4}\lambda_2\Lambda_4\chi_2^2\eta_2^2\right. \\ & & \left.\left.+\dfrac{1}{2}\lambda_2\Lambda_4\chi_2^2{\phi_2^c}^2+\dfrac{1}{2}\lambda_2\Lambda_4\eta_2^2{\phi_2^c}^2\right)\right],
\end{array}
\end{equation}
\\
For the cosmological evolution of the fields $\eta_2$, $\chi_2$ and $\phi_2^c$, the above equation of motion are solved with the Friedmann equations numerically in the flat Universe ($\kappa=0$). The initial conditions used are ${\eta_2}_{ini}=M_P$, ${\chi_2}_{ini}=M_P$, ${\phi_2^c}_{ini}=0$, ${\dot{\eta}_2}_{ini}=0$, ${\dot{\chi}_2}_{ini}=0$ and ${\dot{\phi}_2^c}_{ini}=0$.

Note that after imposing the $Z_2$ symmetry there are five parameters which determine the masses of Higgs fields. With $m_{\eta_1}=m_{H_{SM}}=125.7$GeV and constraint from phase transitions, we have only two equations to determine parameters values\footnote{Constraint on parameters coming from tree level MSSM have not been imposed.}. The Higgs fields masses in this analysis were calculated using eq. (\ref{inertvacuumHiggsmasses}). After imposing $m_{\eta_1}=m_{H_{SM}}=125.7$GeV and constraint from phase transitions eq. (\ref{phasetransitionsconstraint}) some the arbitrary choice of parameters was taken\footnote{The charged Higgs mass was still chosen 
	greater than $80$GeV as suggested by Particle date group (PDG) \cite{PDG}.} since there were only two equations to determine five unknowns.
\\
Masses of Higgs bosons in the analysis are taken to be
\begin{equation*}
m_{\eta_2}=6.925\times 10^{-61}\text{GeV,\quad} m_{\chi_2}=6.925\times 10^{-61}\text{GeV,\quad} m_{\phi_2^\pm}=154.305\text{GeV},
\end{equation*}
just to set the evolution of the energy densities as observed. But too small masses of lightest Higgs fields were required to induce high damping in the field's oscillations which is required to ensure that $\omega_{\text{eff}}$ get less than $-1/3$ only once on the cosmological scale and whenever it gets it never gets value greater than $-1/3$. Since the vacuum energy of the Higgs is too small $\mathcal{O}(10^{-121}E_P/L_P^3)$, it asks for very small values of parameters to achieve over-damped oscillations. 
Thus, the fine tuning problem related to the smallness of field mass in scalar field models also exists in our model too.

To make the Universe evolving in the vacuum states given by Inert vacuum II, $m_{22}^2=0$ has been chosen which implies $T_{c_2}=0$. In this situation, phase transitions in the second doublet will never occur (no appearance of odd forces and particles would be observed) and the Universe will keep on accelerating forever.

The solution of the eqs. (\ref{etaequationofmotion}, \ref{chiequationofmotion}, \ref{chargedHiggsequationofmotion}) along with Friedmann equations is shown below
\begin{figure}[H]
	\centering
	\includegraphics[scale=.25]{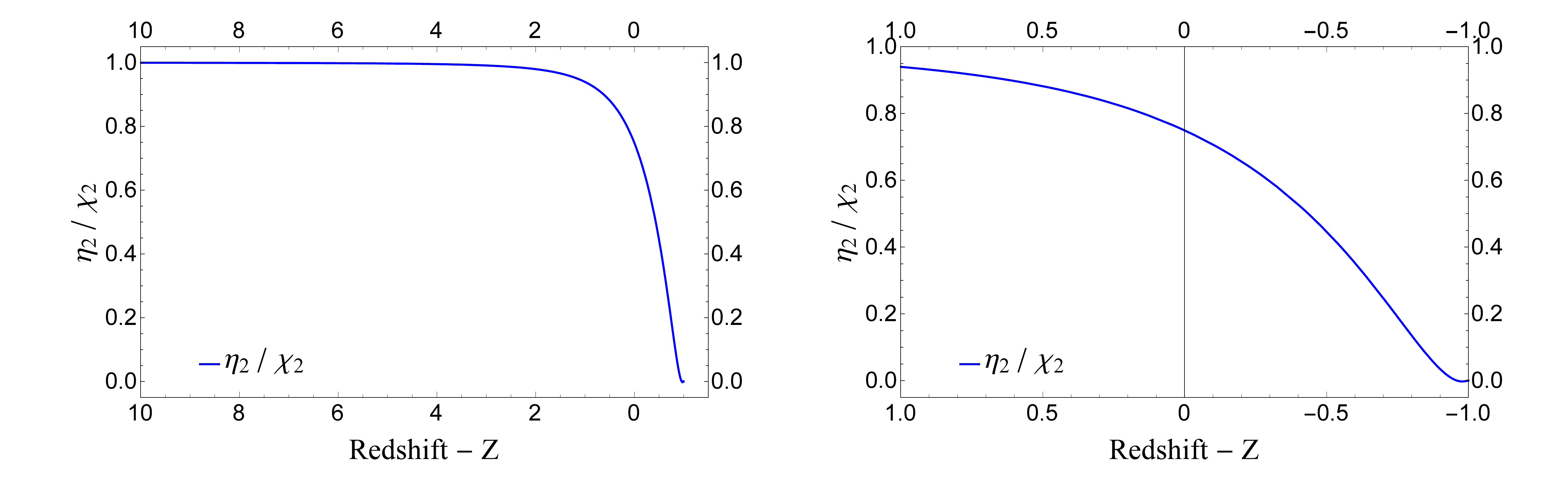}
	\caption{Higgs field as a function of redshift.
	}
	\label{fig:field}
\end{figure}
\begin{figure}[H]
	\centering
	\includegraphics[scale=.25]{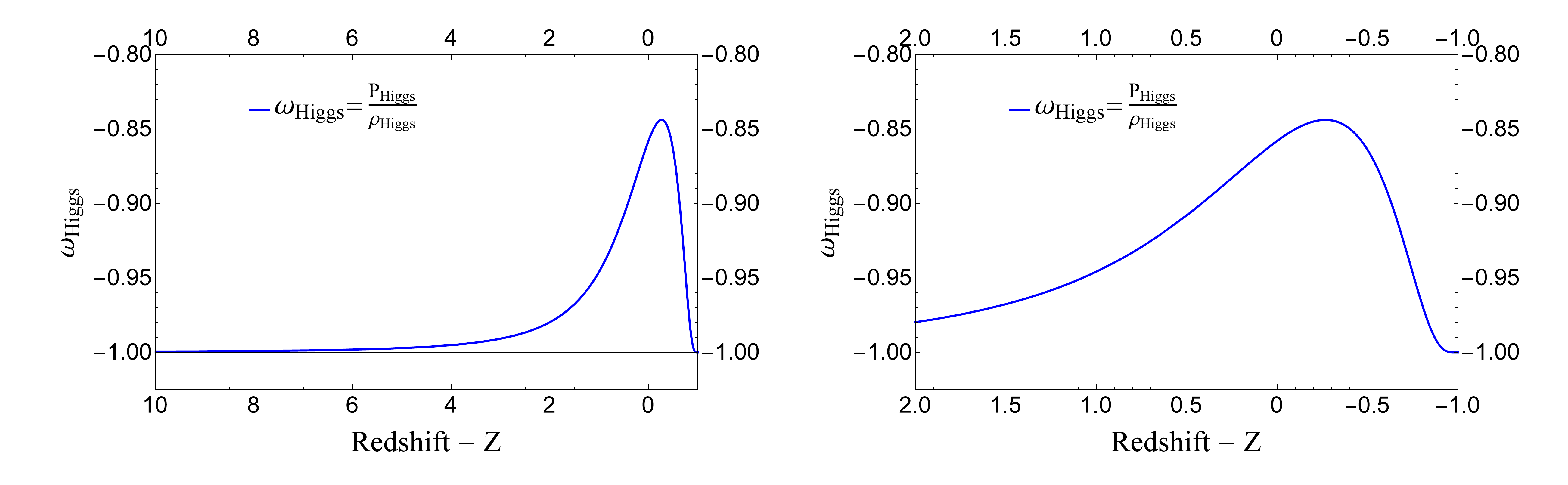}
	\caption{Equation of state $\omega_{Higgs}$ for Higgs fields, as seen it starts with $-1$ then evolves towards quintessence regime after large enough time it comes back at $-1$.}
	\label{fig:OmegaHiggs}
\end{figure}
During the initial stages $Z\approx3750$, the evolution of the Higgs fields $\eta_2$, $\chi_2$ and $\phi_2^{\pm}$ are frozen by the expansion shown in fig. (\ref{fig:field}) and acts as a negligibly small vacuum energy component with $\omega=-1$. As time proceeds, Higgs fields begins to evolve towards minimum of potential, the energy density in the Higgs fields start to dominate cosmologically (on Hubble scale). During the evolution equation of state parameter $\omega_{Higgs}$ starts to increase and becomes $>-1$ as shown in fig. (\ref{fig:OmegaHiggs}). In the very late (future) Universe ($Z\leq0$), the fields comes to rest at the minimum of the potential and a period with $\omega=-1$ is reachieved to give an exponentially accelerating Universe. Since, $\omega_{Higgs}\ngtr-1/3$ at any time in evolution, after $\omega_{\text{eff}}$ gets less than $-1/3$ in the evolution, an ever accelerating Universe is obtained in this model.

\begin{figure}[H]
	\centering
	\includegraphics[scale=.25]{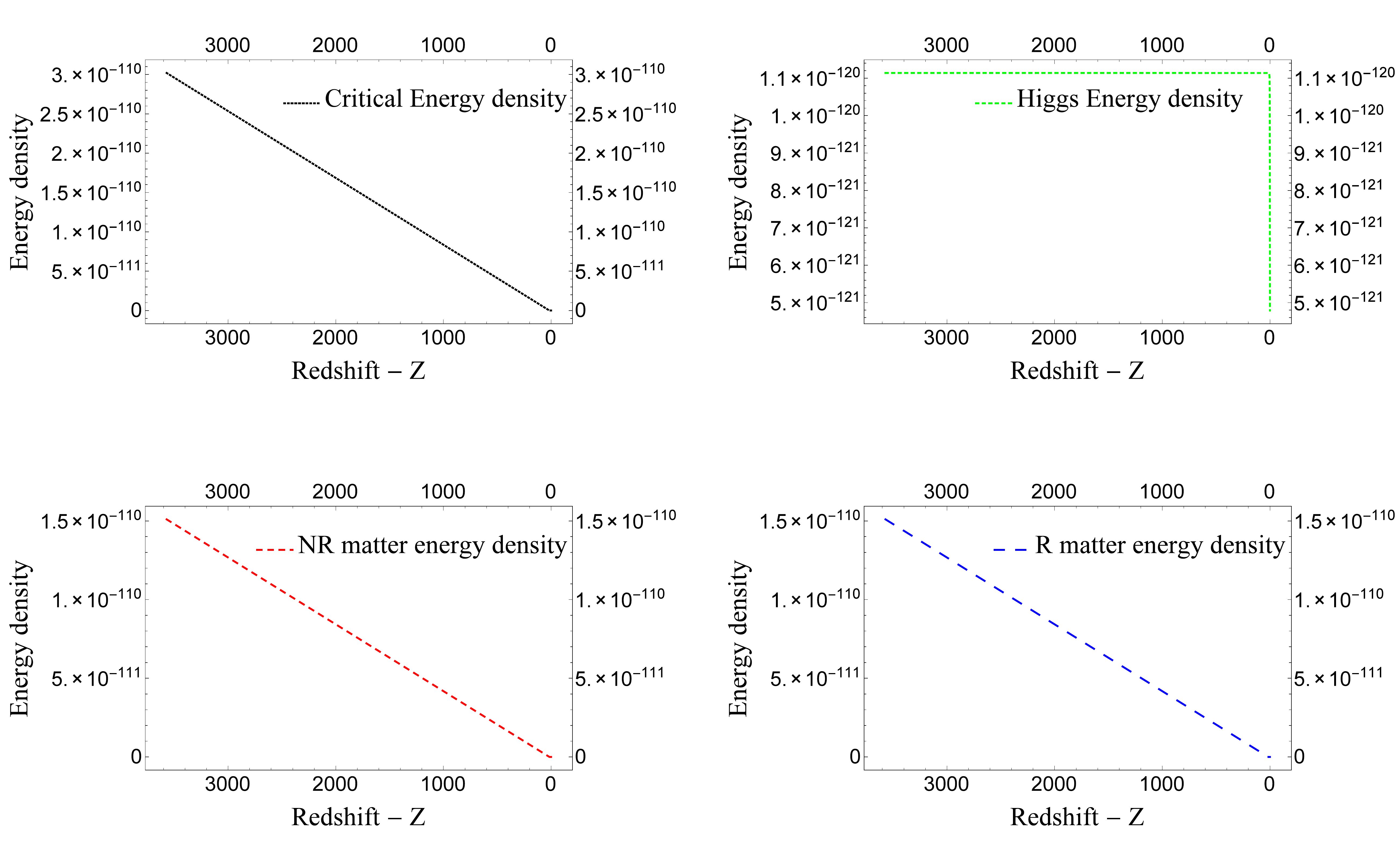}
	\caption{Energy densities of different components as a function of redshift.}
	\label{fig:energydensities1}
\end{figure}
\begin{figure}[H]
	\centering
	\includegraphics[scale=0.5]{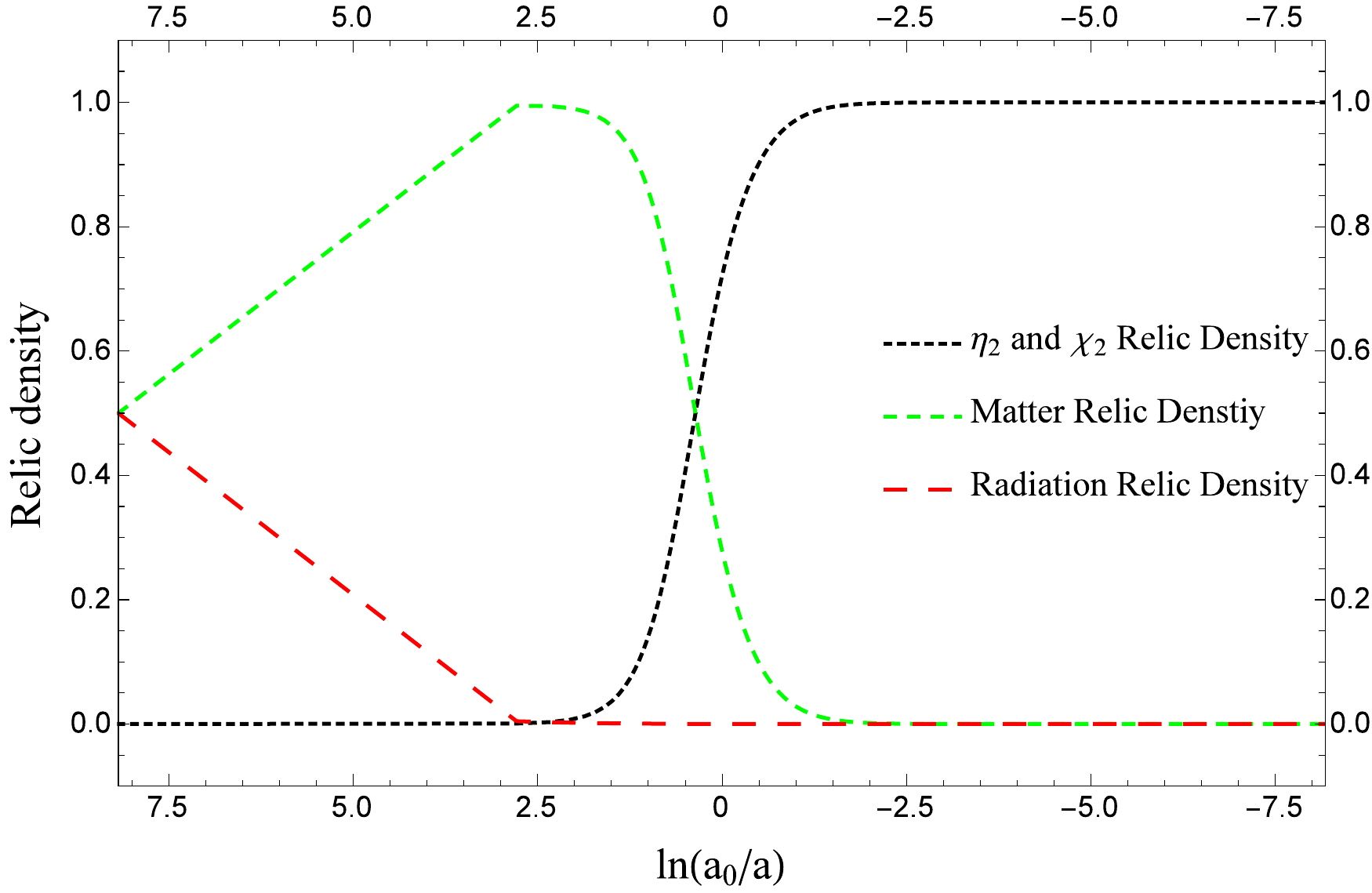}
	\caption{Relic densities of different components as a function of $ln$[$a_0/a$].}
	\label{fig:relicdensities1}
\end{figure}
\begin{figure}[H]
	\centering
	\includegraphics[scale=0.5]{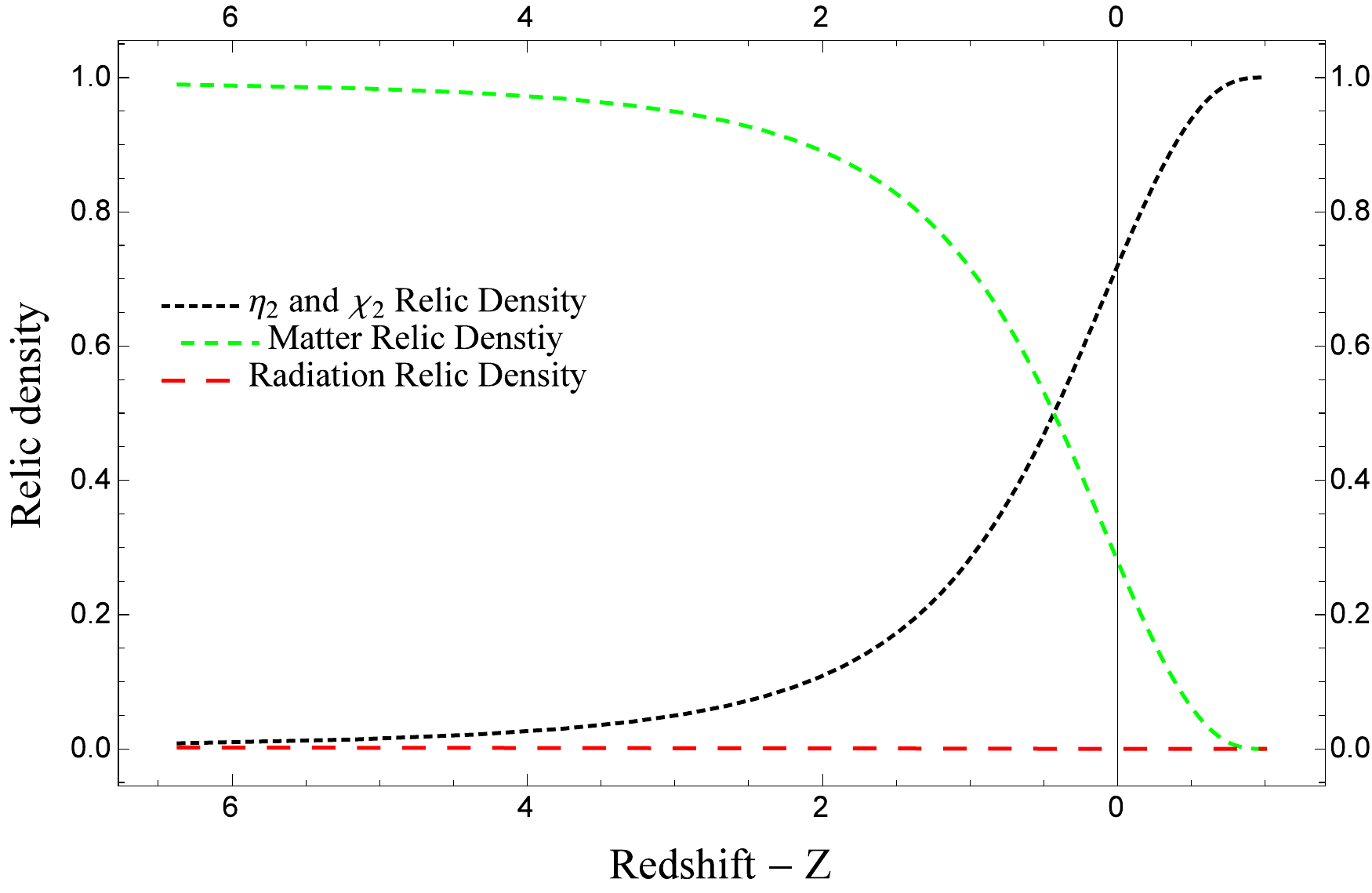}
	\caption{Relic densities of different components as a function of redshift.}
	\label{fig:relicdensities2}
\end{figure}

As discussed before, Higgs field stability is needed to get an ever accelerating expanding Universe. This was obtained by imposing the discrete $Z_2$ symmetry. The lightest Higgs fields $\eta_2$ and $\chi_2$ in this case do not decay into any other Higgs field (since these fields are lighter than SM like and charged Higgs) and also into fermions (since they do not couple to them). Thus, we have a model of ever accelerating expanding Universe.
\begin{figure}[H]
	\centering
	\includegraphics[scale=.25]{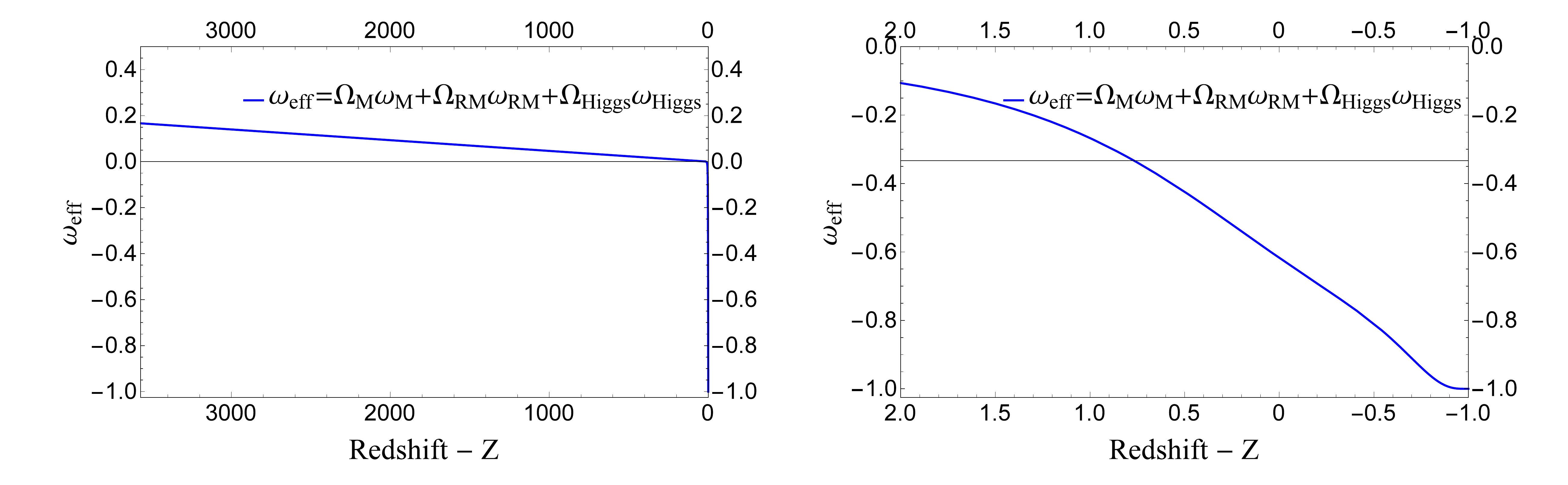}
	\caption{Effective equation of state $\omega_{eff}=\Omega_{DE}\omega_{DE}+\Omega_{R}\omega_R+\Omega_{M}\omega_M$.
	}
	\label{fig:Omegaeffective}
\end{figure}
The $\omega_{\text{eff}}$ in the fig. (\ref{fig:Omegaeffective}) starts from $\approx0.167$ (set by initial conditions $\Omega_{{Higgs}_{int}}=0$ and $\Omega_{{NR}_{int}}=\Omega_{R_{int}}=0.5$; NR: non-relativistic matter and R: relativistic matter) and decreases as the relativistic matter's energy density decreases. This has been shown in fig. (\ref{fig:Omegaeffective}) for $\omega_{\text{eff}}$ and in fig. (\ref{fig:energydensities1}) for energy densities and comes down at $0$. This and before is the time period when non-relativistic matter dominates and expanding Universe decelerates with highest rate ($\Omega_{NR}\approx1$) as non-relativistic matter domination pulls every thing inwards more than outwards Higgs negative pressure. After that $\omega_{\text{eff}}$ starts to decrease as the non-relativistic energy density decreases and Higgs relic energy density increases as shown in fig. (\ref{fig:relicdensities1}) and (\ref{fig:relicdensities2}) this time and afterwards Higgs negative pressure starts to dominates for forever (because second phase transition never occurs in our model) and $\omega_{\text{eff}}$ eventually settles down at $-1$.	
\begin{figure}[H]
	\centering
	\includegraphics[scale=.25]{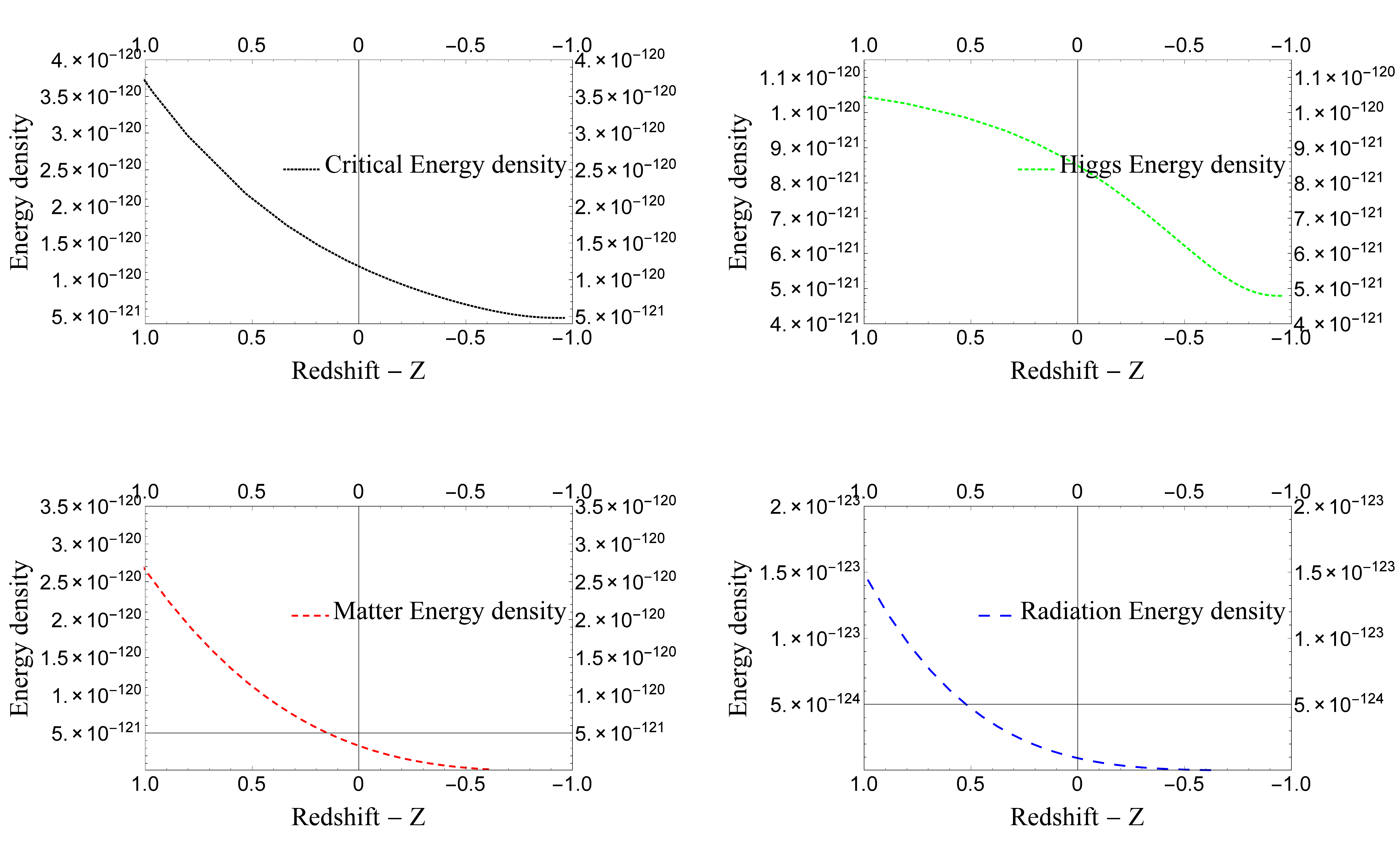}
	\caption{Energy densities of different components as a function of redshift.}
	\label{fig:energydensities2}
\end{figure}
Figure (\ref{fig:energydensities1}) and (\ref{fig:energydensities2}) show the evolution of different components of the Universe as a function of redshift $Z$. In Fig. (\ref{fig:energydensities1}) we can see the evolution of different components from $Z\approx 3750$ to $Z\approx -1$ while fig. (\ref{fig:energydensities2}) shows the evolution from $Z=1$ to $Z\approx -1$. From fig. (\ref{fig:energydensities2}), we note that the energy density decreases as a function of $a^{-3}$ ($a^{-4}$) for non-relativistic matter (relativistic matter) while from fig. (\ref{fig:energydensities1}) it can be seen that for high redshifts the energy density for Higgs fields remains approximately constant. However, for low redshifts (from fig. (\ref{fig:energydensities2})) the Higgs energy density decreases to settle down at their minimum value ($E_{vac}$ given by eq. ({\ref{Evac}})) as the fields come to rest at low redshifts (shown in fig. (\ref{fig:field})).

Figure (\ref{fig:relicdensities1}) is plotted for the evolution of the relic density of the non-relativistic matter, relativistic matter and the dark energy Higgs fields against $ln(a_0/a(t)),$ starting from the non-relativistic and relativistic matter equality; where $t$ is time. The figure shows that only now the dark energy Higgs has started to dominate in the critical density while in the past non-relativistic matter was dominating.
Figure (\ref{fig:relicdensities2}) is plotted for the evolution of the relic density of the non-relativistic matter, relativistic matter and the dark energy Higgs fields against the redshift $Z$ starting from the non-relativistic and relativistic matter equality to $Z\approx-1$; where $t$ is time. The figure shows that for $Z\leq6$ the dark energy Higgs has started to become prominent. At $Z\approx 0.3235$ the non-relativistic matter and relativistic matter energy densities are equal. For $Z > 0.3235$ non-relativistic matter is dominating and for $Z < 0.3235$ dark energy Higgs started to dominate causing the accelerated expansion of the Universe.

It is quite clear from the plots that component Higgs fields of second Higgs doublet are a possible candidate for the dark energy if the present Universe happens to be evolving in the vacuum states described by inert doublet model given by the vacuum solution eq. (\ref{inertvacuum2}).
\\
One needs to know that the initial conditions for charged field were taken in accordance with the observation that dark energy (vacuum) is not charged.
\section{Decay(s) of Higgs field(s)}
At tree level the doublet $\phi_2$ in inert doublet model discussed above whose component fields are the candidate for dark energy is only coupled to the doublet $\phi_1$. The doublet $\phi_1$ acts in an identical way as SM Higgs doublet.
\\
The interaction Lagrangian of Higgs fields of doublet $\phi_2$ with Higgs fields of doublet $\phi_1$ and gauge bosons (extracted from eq. (\ref{LH})) is
\begin{equation}\begin{array}{rcl}\label{Lint}
\mathscr{L}_I &=& \dfrac{1}{2}\nu_1(\lambda_3+\lambda_4-\lambda_5)\chi_2{}^2\eta_1+\dfrac{1}{2}\nu_1(\lambda_3+\lambda_4+\lambda_5)\eta_2{}^2\eta_1+\lambda_3\nu_1\eta_1\phi_2^+\phi_2^-
\vspace{0.2cm} \\ &&
+\dfrac{1}{4}(\lambda_3+\lambda_4-\lambda_5)\chi_2{}^2\eta_1{}^2+\dfrac{1}{4}(\lambda_3+\lambda_4+\lambda_5)\eta_2{}^2\eta_1{}^2+\dfrac{1}{2}\lambda_3\eta_1{}^2\phi_2^+\phi_2^-
\vspace{0.2cm} \\ &&
+\dfrac{g_2^2g_2'{}^2}{(g_2'{}^2+g_2^2)}\phi_2^+\phi_2^-A_{\mu}^2 +\dfrac{(g_2'{}^2+g_2^2)}{8}\eta_2^2Z_{\mu}^2+\dfrac{(g_2'{}^2+g_2^2)}{8}\chi_2^2Z_{\mu}^2+\dfrac{(g_2^2-g_2'{}^2){}^2}{4(g_2'{}^2+g_2^2)}\phi_2^+\phi_2^-Z_{\mu }^2
\vspace{0.2cm} \\ &&
+\dfrac{g_2g_2'(g_2^2-g_2'{}^2)}{(g_2'{}^2+g_2^2)}\phi_2^+\phi_2^-A_{\mu}Z_{\mu}+\dfrac{g_2^2}{4}W^-{}_\mu W^+{}_\mu(\eta_2^2+\chi_2^2+2\phi_2^+\phi_2^-)
\vspace{0.2cm} \\ &&
+\dfrac{g_2{}^2g_2'}{2\sqrt{(g_2'{}^2+g_2^2)}}\eta_2A_{\mu}(\phi_2^+W^-{}_\mu+\phi_2^-W^+{}_\mu)+i\dfrac{g_2{}^2g_2'}{2\sqrt{(g_2'{}^2+g_2^2)}}\chi_2A_{\mu}(\phi_2^-W^+{}_\mu-\phi_2^+W^-{}_\mu)
\vspace{0.2cm} \\ &&
-\dfrac{g_2g_2'{}^2}{2\sqrt{(g_2'{}^2+g_2^2)}}\eta_2Z_\mu(\phi_2^+W^-{}_\mu+\phi_2^-W^+{}_\mu)-i\dfrac{g_2g_2'{}^2}{2\sqrt{(g_2'{}^2+g_2^2)}}\chi_2Z_\mu(\phi_2^-W^+{}_\mu-\phi_2^+W^-{}_\mu).
\end{array}
\end{equation}
To suppress the interaction of Higgs fields $\eta_2$, $\chi_2$ and $\phi_2^\pm$ with the gauge bosons the idea is that the SU(2) doublet $\phi_2$ is very weakly (and different than $\phi_1$) coupled with the gauge bosons, thus here $g_2\lll g_1\text{ and }g_2'\lll g_1'$. Thus, the decay modes that include the Higgs fields of the second Higgs SU(2) doublet are negligible as compared to the modes that include the SM Higgs.
\\
The Higgs decay to a pair of (Higgs) scalar while considering only the on-shell width
, the decay width is given by \cite{2hdmc}
\begin{equation}\label{eta1eta2eta2}
\Gamma(H_i\longrightarrow H_j H_k) = (2-\delta_{jk})m_{H_i}\dfrac{|C_{H_i H_j H_k}|^2}{32\pi}\sqrt{f\left(1,\dfrac{m_{H_j}^2}{m_{H_i}^2},\dfrac{m_{H_k}^2}{m_{H_i}^2}\right)}~~,
\end{equation}
where
\begin{equation*}\label{function}
f\left(1,\dfrac{m_{H_j}^2}{m_{H_i}^2},\dfrac{m_{H_k}^2}{m_{H_i}^2}\right)=\left(1-\dfrac{m_{H_j}^2}{m_{H_i}^2}-\dfrac{m_{H_k}^2}{m_{H_i}^2}\right)^2-4~\dfrac{m_{H_j}^2}{m_{H_i}^2}~\dfrac{m_{H_k}^2}{m_{H_i}^2}~.
\end{equation*}
$C_{H_i H_j H_k}$ is the coupling of different Higgs bosons $H_i$, $H_j$ and $H_k$.
\\
From eq. (\ref{Lint}) $C_{\eta_1 \chi_2 \chi_2}=\dfrac{1}{2}\nu_1(\lambda_3+\lambda_4-\lambda_5)=\dfrac{m_{\chi_2}^2}{\nu_1}$, $C_{\eta_1 \eta_2 \eta_2}=\dfrac{1}{2}\nu_1(\lambda_3+\lambda_4+\lambda_5)=\dfrac{m_{\eta_2}^2}{\nu_1}$ and $C_{\eta_1 \phi_2^+ \phi_2^-}=\lambda_3\nu_1=2\dfrac{{m_{\phi_2^c}^2}}{\nu_1}$. The decay rate of $\eta_1$ to $\eta_2\eta_2$, $\chi_2\chi_2$, $\phi_2^-\phi_2^+$ for the masses used in the cosmological evolution determination in the previous section are,
\begin{equation*}\begin{array}{rcl}\label{decayratesformasses}
&	\Gamma(\eta_1 \longrightarrow \chi_2 \chi_2) &= 1.53794\times 10^{-6}\text{ GeV}
\vspace{0.2cm} \\ &
\Gamma(\eta_1 \longrightarrow \eta_2 \eta_2) &= 1.53794\times 10^{-6}\text{ GeV}
\vspace{0.2cm} \\ &
\Gamma(\eta_1 \longrightarrow \phi_2^+ \phi_2^-) &= 0.\text{ GeV}
\vspace{0.2cm}
\end{array}
\end{equation*}
Also the decay rates for SM like Higgs boson is given in the graph below as a function of masses $m_{\eta_2}$, $m_{\chi_2}$ and $m_{\phi_2^\pm}$,
\begin{figure}[H]
	\centering
	\includegraphics[scale=0.6]{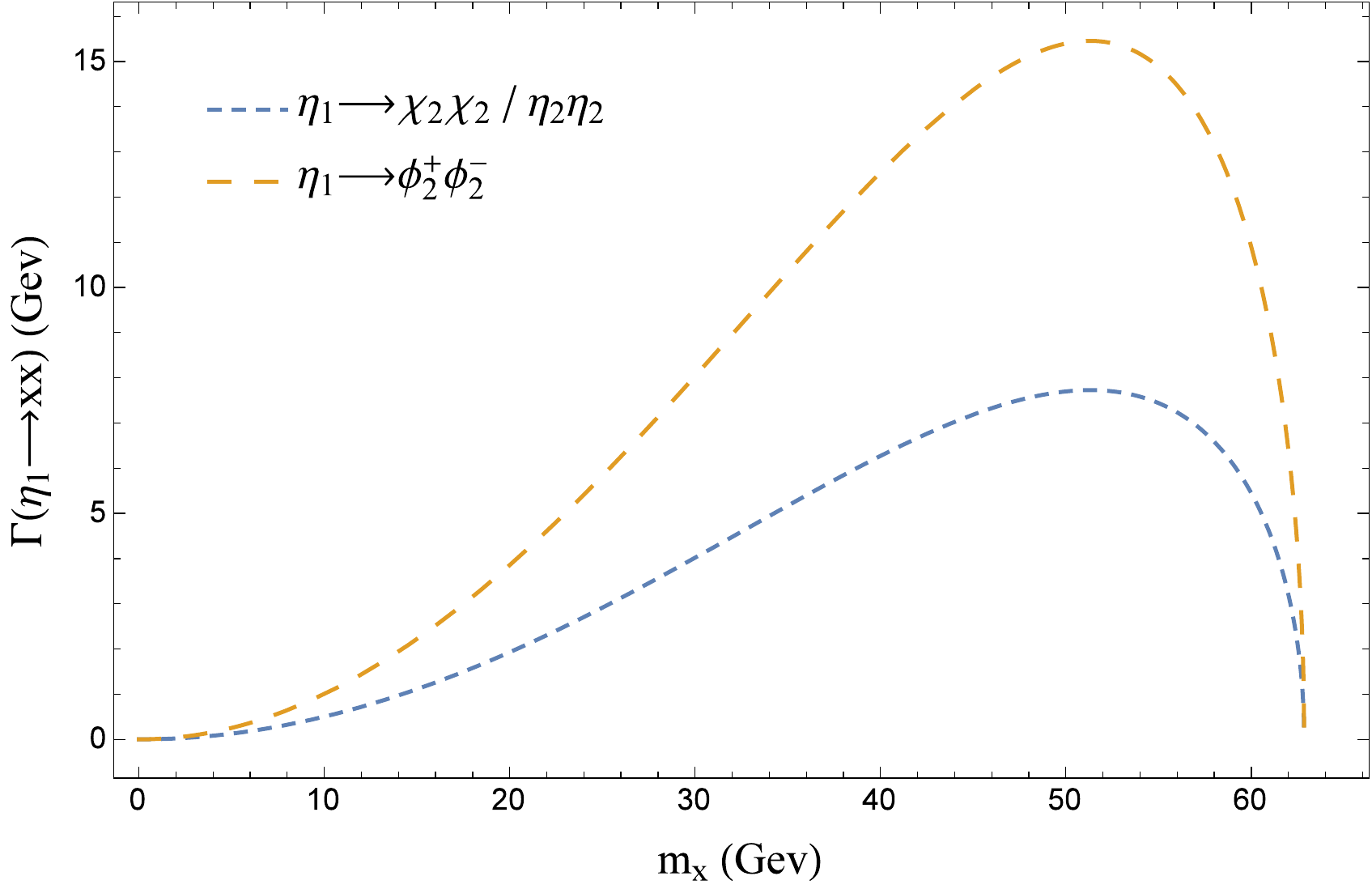}
	\caption{Decay rates for $\eta_1\longrightarrow$ xx}
	\label{fig:decayrates}
\end{figure}
From the above plot, we see that the decay rate for SM like Higgs to other Higgs pairs is peaked at $\text{m}_\text{x}=51.3168$GeV. For $\text{m}_\text{x}<51.3168$GeV the decay rate increases from zero to $\Gamma=15.4557$GeV and $\Gamma=7.72786$GeV for charged pair and neutral pair respectively, after that it decreases sharply and goes to zero at $\text{m}_\text{x}=62.85$GeV.

It should also be mentioned that total decay width of SM like Higgs boson here is well within bounds of mass resolution $\approx12\times10^{-3}$GeV of 
LHC \cite{PDG} (this is only the current results and the uncertainty in the measurements is high \cite{PDG} which is expected to be refined [could come to SM expected value as more data is collected]) for the masses used in the cosmological evolution of the Higgs fields. One should also mention that the SM prediction of total decay width for Higgs boson is $4.21\times10^{-3}$GeV with mass $126$GeV \cite{LHC-CSWG} and is $4.07\times10^{-3}$GeV with mass $125$GeV \cite{PDG}. In this model, we get three more decay channels of SM like Higgs which are to the other Higgs bosons pair.

It should also be mentioned that the dark energy density does not modify as a consequence of the SM-like Higgs decay into dark energy Higgs pair for the masses of the Higgs fields taken here. Had we been able to achieve accelerated expansion with the mass of the second Higgs $\geq125\text{GeV}$  (which we could not because of fine tuning problem) the dark energy density would change significantly at redshift corresponding to the equilibrium temperature (freeze out temperature of dark energy Higgs) $T_{eq}=m_{H_2}-2m_{H_{SM}}$. In that case, we could have a Universe whose different regions would accelerate, decelerate or contract differently $-$ a case usually called bubble Universes, multi-verses or parallel Universes. 
These decays do not imply any additional fine tuning.
\section{Conclusions}\label{Conclusion}
Scalar fields are among the possible candidate for the observed accelerated expansion of the Universe. In this article, I argued that the Particle Physics developed so far must have something in/beyond SM which will explain the observed accelerated expansion of the Universe and hence will serve as the dark energy candidate. This is also necessary for the unification of matter and vacuum energy. This unification is necessary in my opinion. This article took specifically this approach that dark energy actually is some form of scalar field(s) which is(are) present in the so called inert doublet model.

I have found that if the present Universe is described by the vacuum given by an inert doublet model then the component scalar fields of the inert doublet $\phi_2$ can be one possible candidate for the dark energy. Since the present contribution of the dark energy to the critical energy density is about $0.7$, this value is obtained by taking the masses of CP-even field(s) very low. The most important thing is that with the initial conditions set the mass of the charged ($\phi_2^c$) field become arbitrary if we release the parametric constraint given by eq.  (\ref{phasetransitionsconstraint}) which is obtained using phase transitions bound, this model will fit for any value of mass of $\phi_2^c$. One needs to keep in mind that the values of masses were chosen arbitrary just to get dark energy relic density $\approx 0.7$, changing the values of masses mildly does not changes the relic density much. We could also have some other masses set which would give the same evolution as shown below for above masses set but there is some bound on the parameters of the potential to get high damping in the fields (which restricts the masses to remain in certain order of energy). 

As discussed above even though the masses were chosen arbitrary to get $\Omega_{Higgs}\approx0.7$, to avoid the oscillations in the fields we have to take $\lambda_3+\lambda_4+\lambda_5\approx\mathcal{O}(10^{-125})$ and $\lambda_3+\lambda_4-\lambda_5\approx\mathcal{O}(10^{-125})$ and thus the masses of second Higgs doublet very small to ensure that we get $\omega_{\text{eff}}=-1/3$ only once in the history of the Universe. The obtained results suggests to probe the low energy behavior at the Universe's current temperature ($T=2.73^{\text{o}}$K) of the standard model and its extensions to obtain new information and possibility of dark energy candidacy in Particle Physics.

To suppress the interaction of the Higgs fields $\eta_2$, $\chi_2$ and $\phi_2^\pm$ with the gauge bosons $W^{\pm}$, $Z$ and $A(\mu)$,  we imposed the condition that the SU(2) doublet $\phi_2$ is very weakly (and different than $\phi_1$) coupled with the gauge bosons. Thus here $g_2\lll g_1\text{ and }g_2'\lll g_1'$. This allowed us to conclude that the decay modes that include the Higgs fields $\eta_2$, $\chi_2$ and $\phi_2^\pm$ are negligible as compared to the modes that include the SM Higgs. The couplings $g_1 \text{ and }g_1'$ have the values of the SM Higgs gauge couplings in the SM but the values of $g_2 \text{ and }g_2'$ must be set using the results from LHC or any other accelerator.

One thing that remains important to check in all extensions of the SM is whether the Higgs potential contains the vacuum instability or not? If it does, then does it make the vacuum instability more worse as compared to the SM or not?  The answer to the question for our model is that although it contains the vacuum instability, due to the coupling of the second Higgs with the SM Higgs is $\mathcal{O}(10^{-126})$, it will not affect the RGEs running of the SM Higgs. Thus will not make the vacuum instability worse. We expect the vacuum instability to occur at approximately the same scale as it occurs in the SM.

When $T$ becomes less than $T_{c_{2}}$, symmetry in both fields is broken and all mixed fields of both doublets get massive (except Goldstone bosons). In this situation, one neutral Higgs can be made to acts in the same way as SM Higgs does (i.e. giving masses to other particles and having mass $125.7\text{GeV}$) while other fields can be made to act as dark energy field(s) by setting the Yukawa interaction to cancel the other fields effect on fermions. This should be investigated in future.

When describing a model for accelerated expansion of the Universe, it becomes essential to compare it with the $\Lambda$CDM model. 
In comparing our model with the $\Lambda$CDM, we note that on the cosmological scale our model is completely different from the standard $\Lambda$CDM and also very predictive. In our model $\omega_{Higgs}$ is not constant over redshift which has been shown in fig. (\ref{fig:OmegaHiggs}). From fig. (\ref{fig:OmegaHiggs}), we see that the model gives $\omega_{Higgs} \approx-0.858$ at $Z=0$. Thus our model can be distinguished from the $\Lambda$CDM via the variation of $\omega_{Higgs}$  from $-1$ on the cosmological scale. The deviation of $\omega_{Higgs}$ at low redshift $2<Z<0$ is given in fig. (\ref{fig:OmegaHiggs}). In the plot, we see that while we started from $\omega_{Higgs}=-1$ at $Z\approx3750$, around $Z\approx 6$ $\omega_{Higgs}$ start to get greater than $-1$. At redshift $Z=0$ the deviation of $\omega_{Higgs}$ from $-1$ we obtain in our model for explicit values of parameters using MATHEMATICA software is $0.1419828\pm 3.2433649325\times10^{-8}$. 
In concluding, since we get $\omega_{eff}<-1/3$ after solving the Euler Lagrange equations numerically, the proposed Higgs field could cause the current observed accelerated expansion.

It is also worth concluding that because of the type of Yukawa interaction the second phase transition (the case of $m_{22}^2\neq 0$) will not change fermions phenomenology but due to the SU(2)$\otimes$U(1) gauge invariance gauge bosons phenomenology will get changed (in general). Setting the coupling constants $g_2$ and $g_2'$ might end up not changing gauge bosons phenomenology subsequently. This should be investigated in future theoretically.
\section*{Acknowledgment}
I thank Prof. Asghar Qadir for suggesting the idea and giving useful comments on the work. I would also like to thank Rizwan Khalid for his useful discussion on the issue of the stability of the vacuum. At last, I thank {\it{The Abdus Salam International centre for Theoretical Physics (AS-ICTP), Trieste, Italy}} for their hospitality in August, 2014 where part of work was done.

This work is supported by {\textit{National University of Sciences and Technology (NUST), Sector H-12 Islamabad 44000, Pakistan}} and \emph{Higher education commission (HEC) of Pakistan} under the project no. NRPU-3053.
\nocite{*}
\bibliography{Muhammad-Usman.bib}

\providecommand{\noopsort}[1]{}\providecommand{\singleletter}[1]{#1}%
\begin{thebibliography}{27}%
\makeatletter
\providecommand \@ifxundefined [1]{%
 \@ifx{#1\undefined}
}%
\providecommand \@ifnum [1]{%
 \ifnum #1\expandafter \@firstoftwo
 \else \expandafter \@secondoftwo
 \fi
}%
\providecommand \@ifx [1]{%
 \ifx #1\expandafter \@firstoftwo
 \else \expandafter \@secondoftwo
 \fi
}%
\providecommand \natexlab [1]{#1}%
\providecommand \enquote  [1]{``#1''}%
\providecommand \bibnamefont  [1]{#1}%
\providecommand \bibfnamefont [1]{#1}%
\providecommand \citenamefont [1]{#1}%
\providecommand \href@noop [0]{\@secondoftwo}%
\providecommand \href [0]{\begingroup \@sanitize@url \@href}%
\providecommand \@href[1]{\@@startlink{#1}\@@href}%
\providecommand \@@href[1]{\endgroup#1\@@endlink}%
\providecommand \@sanitize@url [0]{\catcode `\\12\catcode `\$12\catcode
  `\&12\catcode `\#12\catcode `\^12\catcode `\_12\catcode `\%12\relax}%
\providecommand \@@startlink[1]{}%
\providecommand \@@endlink[0]{}%
\providecommand \url  [0]{\begingroup\@sanitize@url \@url }%
\providecommand \@url [1]{\endgroup\@href {#1}{\urlprefix }}%
\providecommand \urlprefix  [0]{URL }%
\providecommand \Eprint [0]{\href }%
\providecommand \doibase [0]{http://dx.doi.org/}%
\providecommand \selectlanguage [0]{\@gobble}%
\providecommand \bibinfo  [0]{\@secondoftwo}%
\providecommand \bibfield  [0]{\@secondoftwo}%
\providecommand \translation [1]{[#1]}%
\providecommand \BibitemOpen [0]{}%
\providecommand \bibitemStop [0]{}%
\providecommand \bibitemNoStop [0]{.\EOS\space}%
\providecommand \EOS [0]{\spacefactor3000\relax}%
\providecommand \BibitemShut  [1]{\csname bibitem#1\endcsname}%
\let\auto@bib@innerbib\@empty
\bibitem [{\citenamefont {Peebles}\ and\ \citenamefont
  {Ratra}(2002)}]{Peebles}%
  \BibitemOpen
  \bibfield  {author} {\bibinfo {author} {\bibfnamefont {P.~J.~E.}\
  \bibnamefont {Peebles}}\ and\ \bibinfo {author} {\bibfnamefont
  {B.}~\bibnamefont {Ratra}},\ }\href@noop {} {} (\bibinfo {year} {2002}),\
  \Eprint {http://arxiv.org/abs/astro-ph/0207347v2} {astro-ph/0207347v2}
  \BibitemShut {NoStop}%
\bibitem [{\citenamefont {Copeland}\ \emph {et~al.}(2006)\citenamefont
  {Copeland}, \citenamefont {Sami},\ and\ \citenamefont
  {Tsujikawa}}]{muhammadsami}%
  \BibitemOpen
  \bibfield  {author} {\bibinfo {author} {\bibfnamefont {E.~J.}\ \bibnamefont
  {Copeland}}, \bibinfo {author} {\bibfnamefont {M.}~\bibnamefont {Sami}}, \
  and\ \bibinfo {author} {\bibfnamefont {S.}~\bibnamefont {Tsujikawa}},\
  }\href@noop {} {} (\bibinfo {year} {2006}),\ \Eprint
  {http://arxiv.org/abs/hep-th/0603057v3} {hep-th/0603057v3} \BibitemShut
  {NoStop}%
\bibitem [{\citenamefont {Sotiriou}\ and\ \citenamefont
  {Faraoni}(2010)}]{Faraoni}%
  \BibitemOpen
  \bibfield  {author} {\bibinfo {author} {\bibfnamefont {T.~P.}\ \bibnamefont
  {Sotiriou}}\ and\ \bibinfo {author} {\bibfnamefont {V.}~\bibnamefont
  {Faraoni}},\ }\href@noop {} {\bibfield  {journal} {\bibinfo  {journal} {Rev.\
  Mod.\ Phys.}\ }\textbf {\bibinfo {volume} {82}},\ \bibinfo {pages} {451}
  (\bibinfo {year} {2010})}\BibitemShut {NoStop}%
\bibitem [{\citenamefont {Nojiri}\ and\ \citenamefont
  {Odintsov}(2011)}]{Nojiri}%
  \BibitemOpen
  \bibfield  {author} {\bibinfo {author} {\bibfnamefont {S.}~\bibnamefont
  {Nojiri}}\ and\ \bibinfo {author} {\bibfnamefont {S.~D.}\ \bibnamefont
  {Odintsov}},\ }\href@noop {} {\bibfield  {journal} {\bibinfo  {journal}
  {Phys.\ Rep.\ 505}\ }\textbf {\bibinfo {volume} {59}} (\bibinfo {year}
  {2011})}\BibitemShut {NoStop}%
\bibitem [{\citenamefont {Carroll}\ \emph {et~al.}(2003)\citenamefont
  {Carroll}, \citenamefont {Hoffman},\ and\ \citenamefont {Trodden}}]{trodden}%
  \BibitemOpen
  \bibfield  {author} {\bibinfo {author} {\bibfnamefont {S.~M.}\ \bibnamefont
  {Carroll}}, \bibinfo {author} {\bibfnamefont {M.}~\bibnamefont {Hoffman}}, \
  and\ \bibinfo {author} {\bibfnamefont {M.}~\bibnamefont {Trodden}},\
  }\href@noop {} {\bibfield  {journal} {\bibinfo  {journal} {Phys.\ Rev.\ D}\
  }\textbf {\bibinfo {volume} {68}},\ \bibinfo {pages} {023509} (\bibinfo
  {year} {2003})}\BibitemShut {NoStop}%
\bibitem [{\citenamefont {Onemli}\ and\ \citenamefont
  {Woodard}(2002)}]{0264-9381-19-17-311}%
  \BibitemOpen
  \bibfield  {author} {\bibinfo {author} {\bibfnamefont {V.~K.}\ \bibnamefont
  {Onemli}}\ and\ \bibinfo {author} {\bibfnamefont {R.~P.}\ \bibnamefont
  {Woodard}},\ }\href@noop {} {\bibfield  {journal} {\bibinfo  {journal}
  {Classical and Quantum Gravity}\ }\textbf {\bibinfo {volume} {19}},\ \bibinfo
  {pages} {4607} (\bibinfo {year} {2002})}\BibitemShut {NoStop}%
\bibitem [{\citenamefont {Onemli}\ and\ \citenamefont
  {Woodard}(2004)}]{PhysRevD.70.107301}%
  \BibitemOpen
  \bibfield  {author} {\bibinfo {author} {\bibfnamefont {V.~K.}\ \bibnamefont
  {Onemli}}\ and\ \bibinfo {author} {\bibfnamefont {R.~P.}\ \bibnamefont
  {Woodard}},\ }\href@noop {} {\bibfield  {journal} {\bibinfo  {journal} {Phys.
  Rev. D}\ }\textbf {\bibinfo {volume} {70}},\ \bibinfo {pages} {107301}
  (\bibinfo {year} {2004})}\BibitemShut {NoStop}%
\bibitem [{Note1()}]{Note1}%
  \BibitemOpen
  \bibinfo {note} {In fact, there is no vector field model to explain the dark
  energy in the literature but A. Golovnev, V. Mukhanov and V. Vanchurin have
  described a vector field model for inflation in \cite {vectorinflation} and
  the same procedure can be used for dark energy too.}\BibitemShut {Stop}%
\bibitem [{\citenamefont {Golovnev}\ \emph {et~al.}(2008)\citenamefont
  {Golovnev}, \citenamefont {Mukhanov},\ and\ \citenamefont
  {Vanchurin}}]{vectorinflation}%
  \BibitemOpen
  \bibfield  {author} {\bibinfo {author} {\bibfnamefont {A.}~\bibnamefont
  {Golovnev}}, \bibinfo {author} {\bibfnamefont {V.}~\bibnamefont {Mukhanov}},
  \ and\ \bibinfo {author} {\bibfnamefont {V.}~\bibnamefont {Vanchurin}},\
  }\href@noop {} {\bibfield  {journal} {\bibinfo  {journal} {JCAP06}\ }\textbf
  {\bibinfo {volume} {009}} (\bibinfo {year} {2008})}\BibitemShut {NoStop}%
\bibitem [{\citenamefont {Guth}(1981)}]{guth}%
  \BibitemOpen
  \bibfield  {author} {\bibinfo {author} {\bibfnamefont {A.~H.}\ \bibnamefont
  {Guth}},\ }\href@noop {} {\bibfield  {journal} {\bibinfo  {journal} {Phys.\
  Rev.\ D}\ }\textbf {\bibinfo {volume} {23}},\ \bibinfo {pages} {347}
  (\bibinfo {year} {1981})}\BibitemShut {NoStop}%
\bibitem [{\citenamefont {Linde}(1983)}]{Linde}%
  \BibitemOpen
  \bibfield  {author} {\bibinfo {author} {\bibfnamefont {A.~D.}\ \bibnamefont
  {Linde}},\ }\href@noop {} {\bibfield  {journal} {\bibinfo  {journal}
  {Physics\ Letters\ B}\ }\textbf {\bibinfo {volume} {129}},\ \bibinfo {pages}
  {177} (\bibinfo {year} {1983})}\BibitemShut {NoStop}%
\bibitem [{\citenamefont {Bassett}\ \emph {et~al.}(2006)\citenamefont
  {Bassett}, \citenamefont {Tsujikawa},\ and\ \citenamefont {Wands}}]{Bassett}%
  \BibitemOpen
  \bibfield  {author} {\bibinfo {author} {\bibfnamefont {B.~A.}\ \bibnamefont
  {Bassett}}, \bibinfo {author} {\bibfnamefont {S.}~\bibnamefont {Tsujikawa}},
  \ and\ \bibinfo {author} {\bibfnamefont {D.}~\bibnamefont {Wands}},\
  }\href@noop {} {\bibfield  {journal} {\bibinfo  {journal} {Rev.\ Mod.\ Phys}\
  }\textbf {\bibinfo {volume} {78}},\ \bibinfo {pages} {537} (\bibinfo {year}
  {2006})}\BibitemShut {NoStop}%
\bibitem [{Note2()}]{Note2}%
  \BibitemOpen
  \bibinfo {note} {This does not mean that it is the only possible unified
  model or this has to be the case but certainly it makes its place in the list
  of dark energy candidacy which has to be checked experimentally.}\BibitemShut
  {Stop}%
\bibitem [{\citenamefont {Greenwood}\ \emph {et~al.}()\citenamefont
  {Greenwood}, \citenamefont {Halstead}, \citenamefont {Poltis},\ and\
  \citenamefont {Stojkovic}}]{PhysRevD.79.103003}%
  \BibitemOpen
  \bibfield  {author} {\bibinfo {author} {\bibfnamefont {E.}~\bibnamefont
  {Greenwood}}, \bibinfo {author} {\bibfnamefont {E.}~\bibnamefont {Halstead}},
  \bibinfo {author} {\bibfnamefont {R.}~\bibnamefont {Poltis}}, \ and\ \bibinfo
  {author} {\bibfnamefont {D.}~\bibnamefont {Stojkovic}},\ }\href@noop {}
  {\bibfield  {journal} {\bibinfo  {journal} {Phys. Rev. D}\ }\textbf {\bibinfo
  {volume} {79}},\ \bibinfo {pages} {103003}}\BibitemShut {NoStop}%
\bibitem [{Note3()}]{Note3}%
  \BibitemOpen
  \bibinfo {note} {\begin {eqnarray}\begin {array}{rcl}\label {Lsm} \protect
  \mathscr {L}^{SM}_{gf} &=& -\protect \frac {1}{4}G_{\mu \nu }G^{\mu \nu
  }-\protect \frac {1}{4}W_{\mu \nu }W^{\mu \nu }-\protect \frac {1}{4}B_{\mu
  \nu }B^{\mu \nu } \\ & & +{\protect \cc@accent {"7016}{\psi
  }}_{L}^{i}\protect \cc@accent {"705F}{\iota }{\gamma ^{\mu }}{{\nabla }_{\mu
  }^{EW}}{{\psi }_{L}^{i}}+{\protect \cc@accent {"7016}{\psi }}_{R}^{i}{\iota
  }{\sigma ^{\mu }}{{\nabla }_{\mu }^{EW}}{{\psi }_{R}^{i}} \\ & & +{\protect
  \cc@accent {"7016}{\chi }}_{L}^{i}{\iota }{\gamma ^{\mu }}{{\nabla }_{\mu
  }^{SM}}{{\chi }_{L}^{i}}+{\protect \cc@accent {"7016}{U}}_{R}^{i}{\iota
  }{\sigma ^{\mu }}{{\nabla }_{\mu }^{SM}}{{U}_{R}^{i}}+{\protect \cc@accent
  {"7016}{D}}_{R}^{i}{\iota }{\sigma ^{\mu }}{{\nabla }_{\mu
  }^{SM}}{{D}_{R}^{i}}~. \end {array} \end {eqnarray}}\BibitemShut {NoStop}%
\bibitem [{Note4()}]{Note4}%
  \BibitemOpen
  \bibinfo {note} {\protect \vspace {-0.5cm} \begin {equation}\begin
  {array}{rcl}\label {Lyukawa} \protect \mathscr {L}_{Y} &=& -Y_{ij}^u{\protect
  \cc@accent {"7016}{\chi }}_L^i\protect \cc@accent {"707E}{\phi
  _1}U_R^j-Y_{ij}^d{\protect \cc@accent {"7016}{\chi }}_L^i{\phi
  _1}D_R^j-Y_{ij}^e{\protect \cc@accent {"7016}{\psi }}_L^i{\phi _1}\psi
  _R^j-h.c. \end {array} \end {equation} here $\psi _L^i$ are left handed
  leptons doublets, $\psi _R^i$ are right handed leptons singlets, $\chi _L^i$
  are left handed quark doublets, $U_R^i$ and $D_R^i$ are the right handed
  quark singlets. $i$ runs from 1-3. $\phi _1$ is the SM like Higgs
  doublet.}\BibitemShut {Stop}%
\bibitem [{Note5()}]{Note5}%
  \BibitemOpen
  \bibinfo {note} {In the SM one scalar isodoublet with hypercharge $Y=1$ is
  sufficient to make the theory complete and gauge invariant and hence in SM
  \begin {equation}\begin {array}{rcl}\label {LSMHiggs} \protect \mathscr
  {L}_{Higgs}=T_{H}-V_{H}=(D_{\mu }\phi )^{\dagger }(D^{\mu }\phi )-(-\protect
  \frac {{\mu }^{2}}{2!}{\phi }^2+\protect \frac {\lambda }{4!}{\phi }^4) \end
  {array} \end {equation} $$\phi =\begin {pmatrix} \phi ^+ \\ \phi ^0 \end
  {pmatrix} \protect \text {\hskip 2em\relax \hskip 2em\relax and\hskip
  2em\relax \hskip 2em\relax } \protect \cc@accent {"707E}{\phi }=\protect
  \cc@accent {"705F}{\iota }\sigma ^2\protect \cc@accent {"7016}{\phi }=\begin
  {pmatrix} \protect \cc@accent {"7016}{\phi }^0 \\ -\protect \cc@accent
  {"7016}{\phi }^+ \end {pmatrix}.$$}\BibitemShut {NoStop}%
\bibitem [{\citenamefont {Ginzburg}\ and\ \citenamefont
  {Krawczyk}(2005)}]{Ginzburg-2}%
  \BibitemOpen
  \bibfield  {author} {\bibinfo {author} {\bibfnamefont {I.~F.}\ \bibnamefont
  {Ginzburg}}\ and\ \bibinfo {author} {\bibfnamefont {M.}~\bibnamefont
  {Krawczyk}},\ }\href@noop {} {\bibfield  {journal} {\bibinfo  {journal}
  {Phys.\ Rev.\ D}\ }\textbf {\bibinfo {volume} {72}},\ \bibinfo {pages}
  {115013} (\bibinfo {year} {2005})}\BibitemShut {NoStop}%
\bibitem [{\citenamefont {Eriksson}\ \emph {et~al.}(2010)\citenamefont
  {Eriksson}, \citenamefont {Rathsman},\ and\ \citenamefont {Stal}}]{2hdmc}%
  \BibitemOpen
  \bibfield  {author} {\bibinfo {author} {\bibfnamefont {D.}~\bibnamefont
  {Eriksson}}, \bibinfo {author} {\bibfnamefont {J.}~\bibnamefont {Rathsman}},
  \ and\ \bibinfo {author} {\bibfnamefont {O.}~\bibnamefont {Stal}},\
  }\href@noop {} {\bibfield  {journal} {\bibinfo  {journal} {Computer\ Physics\
  Communications}\ }\textbf {\bibinfo {volume} {181}},\ \bibinfo {pages} {189}
  (\bibinfo {year} {2010})}\BibitemShut {NoStop}%
\bibitem [{\citenamefont {Kaffas}\ \emph {et~al.}(2007)\citenamefont {Kaffas},
  \citenamefont {Khater}, \citenamefont {Ogreid},\ and\ \citenamefont
  {Osland}}]{kaffaskhater}%
  \BibitemOpen
  \bibfield  {author} {\bibinfo {author} {\bibfnamefont {A.~E.}\ \bibnamefont
  {Kaffas}}, \bibinfo {author} {\bibfnamefont {W.}~\bibnamefont {Khater}},
  \bibinfo {author} {\bibfnamefont {O.}~\bibnamefont {Ogreid}}, \ and\ \bibinfo
  {author} {\bibfnamefont {P.}~\bibnamefont {Osland}},\ }\href@noop {}
  {\bibfield  {journal} {\bibinfo  {journal} {Nucl.\ Phys.\ B}\ }\textbf
  {\bibinfo {volume} {775}},\ \bibinfo {pages} {45} (\bibinfo {year}
  {2007})}\BibitemShut {NoStop}%
\bibitem [{\citenamefont {Ginzburg}\ \emph {et~al.}(2010)\citenamefont
  {Ginzburg}, \citenamefont {Kanishev}, \citenamefont {Krawczyk},\ and\
  \citenamefont {Sokolowska}}]{Ginzburg-1}%
  \BibitemOpen
  \bibfield  {author} {\bibinfo {author} {\bibfnamefont {I.~F.}\ \bibnamefont
  {Ginzburg}}, \bibinfo {author} {\bibfnamefont {K.~A.}\ \bibnamefont
  {Kanishev}}, \bibinfo {author} {\bibfnamefont {M.}~\bibnamefont {Krawczyk}},
  \ and\ \bibinfo {author} {\bibfnamefont {D.}~\bibnamefont {Sokolowska}},\
  }\href@noop {} {\bibfield  {journal} {\bibinfo  {journal} {Phys.\ Rev.\ D}\
  }\textbf {\bibinfo {volume} {82}},\ \bibinfo {pages} {123533} (\bibinfo
  {year} {2010})}\BibitemShut {NoStop}%
\bibitem [{\citenamefont {Branco}\ \emph {et~al.}(2012)\citenamefont {Branco},
  \citenamefont {Ferreira}, \citenamefont {Lavoura}, \citenamefont {Rebelo},
  \citenamefont {Sher},\ and\ \citenamefont {Silva}}]{Brancoetal}%
  \BibitemOpen
  \bibfield  {author} {\bibinfo {author} {\bibfnamefont {G.}~\bibnamefont
  {Branco}}, \bibinfo {author} {\bibfnamefont {P.}~\bibnamefont {Ferreira}},
  \bibinfo {author} {\bibfnamefont {L.}~\bibnamefont {Lavoura}}, \bibinfo
  {author} {\bibfnamefont {M.}~\bibnamefont {Rebelo}}, \bibinfo {author}
  {\bibfnamefont {M.}~\bibnamefont {Sher}}, \ and\ \bibinfo {author}
  {\bibfnamefont {J.~P.}\ \bibnamefont {Silva}},\ }\href@noop {} {\bibfield
  {journal} {\bibinfo  {journal} {Physics\ Reports}\ }\textbf {\bibinfo
  {volume} {516}},\ \bibinfo {pages} {1} (\bibinfo {year} {2012})}\BibitemShut
  {NoStop}%
\bibitem [{\citenamefont {Kapusta}\ and\ \citenamefont {Gale}(2006)}]{kapusta}%
  \BibitemOpen
  \bibfield  {author} {\bibinfo {author} {\bibfnamefont {J.~I.}\ \bibnamefont
  {Kapusta}}\ and\ \bibinfo {author} {\bibfnamefont {C.}~\bibnamefont {Gale}},\
  }\href@noop {} {\emph {\bibinfo {title} {Finite temperature field theory,
  principles and applications}}}\ (\bibinfo  {publisher} {Cambridge University
  Press},\ \bibinfo {year} {2006})\BibitemShut {NoStop}%
\bibitem [{Note6()}]{Note6}%
  \BibitemOpen
  \bibinfo {note} {Constraint on parameters coming from tree level MSSM have
  not been imposed.}\BibitemShut {Stop}%
\bibitem [{Note7()}]{Note7}%
  \BibitemOpen
  \bibinfo {note} {The charged Higgs mass was still chosen greater than $80$GeV
  as suggested by Particle date group (PDG) \cite {PDG}.}\BibitemShut {Stop}%
\bibitem [{\citenamefont {{K. A. Olive et al}}(2014)}]{PDG}%
  \BibitemOpen
  \bibfield  {author} {\bibinfo {author} {\bibnamefont {{K. A. Olive et al}}},\
  }\href@noop {} {\bibfield  {journal} {\bibinfo  {journal} {Chinese\ Physics\
  C}\ }\textbf {\bibinfo {volume} {38}},\ \bibinfo {pages} {090001} (\bibinfo
  {year} {2014})}\BibitemShut {NoStop}%
\bibitem [{\citenamefont {{LHC Higgs Cross Section Working Group: S. Dittmaier
  et al}}(2012)}]{LHC-CSWG}%
  \BibitemOpen
  \bibfield  {author} {\bibinfo {author} {\bibnamefont {{LHC Higgs Cross
  Section Working Group: S. Dittmaier et al}}},\ }\href@noop {} {\emph
  {\bibinfo {title} {Handbook of LHC Higgs Cross Sections: 2. Differential
  Distributions}}},\ \bibinfo {type} {Tech. Rep.}\ (\bibinfo {year} {2012})\
  \bibinfo {note} {{CERN Report 2} (Tables A.1 -– A.20)},\ \Eprint
  {http://arxiv.org/abs/hep-ph/0207347v2} {hep-ph/0207347v2} \BibitemShut
  {NoStop}%
\end{thebibliography}%
\end{document}